\documentclass[preprintnumbers, floatfix, preprintnumbers, letterpaper, superscriptaddress,nofootinbib]{revtex4-2}
\usepackage{graphicx}
\usepackage{microtype}
\usepackage{amsmath}
\usepackage{amssymb}
\usepackage{subfigure}
\usepackage{hyperref}
\usepackage{url}
\usepackage{xcolor}
\usepackage{color}
\usepackage{mathrsfs}
\usepackage{calrsfs}
\usepackage{amsfonts}
\usepackage{lipsum}
\usepackage{eufrak}
\usepackage{tabularx}
\usepackage{eucal}
\usepackage{latexsym}
\usepackage{ragged2e}
\usepackage{epsfig}
\usepackage{textcomp}
\usepackage{orcidlink}
\usepackage{caption}
\DeclareCaptionJustification{justified}{\leftskip=0pt \rightskip=0pt \parfillskip=0pt plus 1fil}
\definecolor{vividviolet}{rgb}{0.62, 0.0, 1.0}
\definecolor{amaranth}{rgb}{0.9, 0.17, 0.31}
\definecolor{palatinateblue}{rgb}{0.15, 0.23, 0.89}
\definecolor{brightpink}{rgb}{1.0, 0.0, 0.5}
\definecolor{cornflowerblue}{rgb}{0.39, 0.58, 0.93}
\definecolor{deepcarminepink}{rgb}{0.94, 0.19, 0.22}
\definecolor{radicalred}{rgb}{1.0, 0.21, 0.37}

\hypersetup{ linktoc=all,
	colorlinks, linkcolor={palatinateblue},
	citecolor={brightpink}, urlcolor={amaranth}
}

\graphicspath{{Images/}}

\def\sideremark#1{\ifvmode\leavevmode\fi\vadjust{\vbox to0pt{\vss
			\hbox to 0pt{\hskip\hsize\hskip1em
				\vbox{\hsize1.3cm\tiny\raggedright\pretolerance10000
					\noindent #1\hfill}\hss}\vbox to8pt{\vfil}\vss}}}%
\def\beq{\begin{equation}}
\def\eeq{\end{equation}}

\begin{document}

\title{On the uniqueness of $\Lambda$CDM-like evolution for homogeneous and isotropic cosmology in General Relativity}

\author{Saikat Chakraborty\orcidlink{0000-0002-5472-304X}}
\email{saikat.ch@nu.ac.th}
\affiliation{The Institute of Fundamental Study ``The Tah Poe Academia Institute'',\\ Naresuan University, Phitsanulok 65000, Thailand}
\affiliation{Center for Space Research, North-West University, Mahikeng 2745, South Africa}

\author{Daniele Gregoris\orcidlink{0000-0002-0448-3447}}
\email{danielegregoris@libero.it}
\affiliation{Department of Physics, School of Science, Jiangsu University of Science and Technology, Zhenjiang 212003, China}

\author{B. Mishra\orcidlink{0000-0001-5527-3565}}
\email{bivu@hyderabad.bits-pilani.ac.in}
\affiliation{Department of Mathematics, Birla Institute of Technology and
Science-Pilani,\\ Hyderabad Campus, Hyderabad-500078, India.}

\begin{abstract}
We address the question of the uniqueness of spatially flat $\Lambda$CDM-like evolution for FLRW cosmologies in General Relativity, i.e. whether any model other than the spatially flat $\Lambda$CDM can give rise to the same type of scale factor evolution. Firstly, we elaborate on what we exactly imply by a $\Lambda$CDM-like evolution or kinematic/cosmographic degeneracy with the $\Lambda$CDM model, using the lessons from the statefinder diagnostic. Then, we  consider two  models with interaction in the dark sector: coupled fluid-fluid model and coupled quintessence model. We enforce the \emph{kinematic} degeneracy with the spatially flat $\Lambda$CDM model via the cosmographic condition $j=1$ ($j$ being the jerk parameter), which in turn fixes the function of the interaction term that is a priori unspecified. We argue that in General Relativity this cosmographic condition is consistent only with spatial flatness.  Employing a dynamical system approach, we show that the spatially flat coupled fluid-fluid interacting models kinematically degenerate with $\Lambda$CDM must necessarily be based on a phantom fluid, whereas the set of physically viable spatially flat coupled quintessence models with power law or exponential potential kinematically degenerate to $\Lambda$CDM is of measure zero. Our analysis establishes that coupled fluid-fluid models with non-phantom fluids or coupled quintessence models with power law and exponential potential can never reproduce a cosmological evolution similar to that of the $\Lambda$CDM. The astrophysical consequences of our findings are qualitatively discussed in light of observational cosmological tensions. 
\end{abstract}

\maketitle

\section{Introduction}
\label{ssintro}

Although  the $\Lambda$CDM model is widely adopted in accounting for various cosmological datasets, now that we have entered an epoch of high precision  cosmological measurements, some tensions have been identified \cite{intertwined1,intertwined2,intertwined22}. Specifically, analyses of CMB \cite{TT1} and of supernovae \cite{TT2} datasets  deliver discrepant estimates of the value of the Hubble constant $H_0$ and, even accounting just for the former  may require a description of dark energy beyond a cosmological constant \cite{TT3}. Moreover, the parameter estimate from the growth of matter perturbations, whose evolution can be reconstructed from redshift space distortion observations \cite{rsd}, challenges that from the Planck and supernovae datasets \cite{tension1,tension2,tension3}. This is the so-called $f \sigma_8$ tension \cite{intertwined3}, where $f$ is the logarithmic growth rate of matter density perturbations and $\sigma_8$ the matter power spectrum at $8h^{-1}$ Mpc. Therefore, we are required to construct some novel theoretical model whose background dynamics resembles that of $\Lambda$CDM as closely as possible, but which also comes with some freedom allowing us to describe a realistic growth of the matter perturbations. The cosmographic approach \cite{cosmot1,cosmot2,snap} and the statefinder diagnostic \cite{state} can be exploited for quantifying the deviations of the cosmological model at hand from $\Lambda$CDM. While model independent estimates of the jerk parameter  relying on cosmic chronometers and supernovae data are consistent with the $\Lambda$CDM condition $j \approx 1$  within $3\sigma$ \cite{reza1,reza2}, the reconstructed cosmographic quantities are in good agreement with those of $\Lambda$CDM when some specific parametric deviations from  the concordance values  are assumed in the analysis   and then constrained by the data fitting \cite{intj1,intj2,intj3,cosmoII}. However, we think that it is necessary to refine the notion of degeneracy between cosmological models by distinguishing their {\it kinematical} vs. {\it dynamical} properties. This requires assessing the different roles of  the underlying evolution (differential) equation {\it and} of the boundary conditions. The redshift evolution of the jerk parameter is (see e.g. \cite[Eq.(4)]{reza1}): 
\beq
j(z)=\frac{H(z) (1+z)^2 \frac{d^2 H(z)}{dz^2}+\left[(1+z) \frac{d H(z)}{dz}-H(z)  \right]^2}{H^2(z)}\,.
\eeq
Solving $j(z)=1$ we obtain \cite{intj1,intj2,intj3,cosmoII} 
\beq
\label{HC1C2}
h^{2}(z)  \equiv \frac{H^2(z)}{H_0^2} = {\mathcal C}_1 (1+z)^3 + {\mathcal C}_2,
\eeq
where ${\mathcal C}_1$ and ${\mathcal C}_2$ are two arbitrary constants satisfying ${\mathcal C}_1 + {\mathcal C}_2=1$. In general, for  nonzero $\mathcal{C}_1$ and $\mathcal{C}_2$, the family of solutions given by Eq.\eqref{HC1C2} specifies a $\Lambda$CDM-like evolution history in the sense that it has two clearly defined limits: an {\it effective} CDM  limit at asymptotic past $z\rightarrow\infty$ (or $a\rightarrow0$) where the cosmic evolution goes like $h^2 \sim (1+z)^3$, and an {\it effective} $\Lambda$-limit at asymptotic future $z\rightarrow-1$ (or $a\rightarrow\infty$) where the evolution goes like $h^2 \sim constant$. The particular solution of the above family, specified by ${\mathcal C}_1 = \Omega_{m0}$ and ${\mathcal C}_2 = \Omega_{\Lambda 0}$, corresponds to the particular $\Lambda$CDM cosmic history that our universe is going through. 

It should be appreciated, though, that the condition $j(z)=1$ by itself does not specify fully the $\Lambda$CDM model. This is best understood in terms of the statefinder parameters $\{r,s\}$ introduced in \cite{state}. The first statefinder parameter $r$ is precisely the cosmographic jerk parameter $j$, but this alone does \emph{not} specify a model. Specification of the model requires specification of both the statefinder parameters $\{r,s\}$. If one puts $r=1$ in \cite[Eq.3]{state}, one sees that the equation can be satisfied in general by many different models, including dynamical dark energy models. However, whatever the inherent model  is, it must give rise to an evolution of the form of Eq.\eqref{HC1C2}, and therefore have clearly defined CDM-limit and $\Lambda$-limit for nonzero integration constants $\mathcal{C}_{1,2}$. Also, note that the condition $j=1$ by itself does \emph{not} provide either any information on the physical meaning of the two integration constants ${\mathcal C}_1$ and ${\mathcal C}_2$, nor any route for fixing them. A physical meaning of these two arbitrary constants can only be provided with respect to a particular model in mind. For example, in the $\Lambda$CDM model, dark matter and dark energy evolve independently, and their energy densities redshift as $(1+z)^3$ and $(1+z)^0$ respectively. Then and only then could we identify ${\mathcal C}_1$ with $\Omega_{m0}$ and ${\mathcal C}_2$ with $\Omega_{\Lambda 0}$. We could not have concluded the same, for example, had the equation of state of dark energy been dynamical or had there been an interaction in the dark sector. In other words, the functional relationships ${\mathcal C}_1 = {\mathcal C}_1 (\Omega_{\rm DE0},\Omega_{m0})$ and ${\mathcal C}_2 = {\mathcal C}_2 (\Omega_{\rm DE0}, \Omega_{m0})$ are sensitive to the choice of the model \cite{intj1}. In what follows we will derive them considering for example a dark energy model  $p_{\rm DE}=w_{\rm DE}\rho_{\rm DE}$ with $w_{\rm DE} \neq -1$ and also allowing for energy interactions between dark energy and dark matter. Therefore, {\it the condition $j=1$ is sufficient for reproducing the kinematics of $\Lambda$CDM (i.e. similar cosmic evolution), but \emph{not} its dynamics (i.e. the model itself)}. From the astrophysical perspective, this  remark means that when testing the cosmic history (\ref{HC1C2}) with respect to cosmographic data, one may very well get ${\mathcal C}_1=0.3$ and ${\mathcal C}_2=0.7$, which however should not be taken naively to represent the abundances of dark matter and dark energy at the present day \cite{intj1}. 

It is worthwhile to mention that, the condition $j(z)=1$ (or $r=1$ in the language of the statefinder diagnostic) always specifies a $\Lambda$CDM-like evolution \eqref{HC1C2} except for the specific cases when $q(z)=\frac{1}{2},-1$, which corresponds to either $\mathcal{C}_2$ or $\mathcal{C}_1$ vanishing in Eq.\eqref{HC1C2}\footnote{Cosmographic parameters $\{q,j\}=\{\frac{1}{2},1\}$ corresponds to statefinder parameters $\{r,s\}=\{1,1\}$, which is the SCDM model in statefinder diagnostic. See \cite[Fig.1]{state}}. In principle, solutions with $q(z)=\frac{1}{2}$ or $q(z)=-1$ also satisfy $j(z)=1$. As we will see in the phase space analysis that we are going to carry out, such solutions represent a zero measure set, i.e. solutions that always reside on an equilibrium point or solutions that can be represented by only one single phase trajectory. Except for these special solutions, for all other solutions the other cosmographic quantities, which can then be used for example for writing the luminosity distance, can be reconstructed as algebraic combinations of the function $H=H(z)$ and its derivatives, and therefore are fully determined by the condition $j(z)=1$ \cite{cosmot1,cosmot2}.
 
 Our intuition at the core of the present paper is that we can have the same kinematics, e.g. cosmographic properties, as in the standard model of cosmology, while also preserving some freedom which may be tuned for alleviating the current observational tensions \cite{intertwined1,intertwined2,intertwined22}. In this paper, we will introduce as well a notion of {\it speciality} for the $\Lambda$CDM-like cosmic evolution. Such an evolution comes with a matter dominated epoch as a past asymptotic and a dark energy dominated epoch as a future asymptotic. In the language of the dynamical system formulation of cosmology, which we adopt in the present paper, the former is an unstable equilibrium or past attractor while the latter is a stable equilibrium or future attractor.  In what follows, we will show that not all the models exhibiting the cosmographic property $j=1$ are viable physical models because they lack a physically viable future attractor. The dynamical evolution makes the system leave the physically viable region as given by the constraints $0 \leq \Omega_m \leq 1$ and  $0 \leq \Omega_{\rm DE} \leq 1$. In particular, we will show that for a coupled fluid cosmology with interactions in the dark sector and satisfying $j=1$ to be physically viable, it should necessarily be based on phantom fluids, while physically viable coupled quintessence cosmologies with exponential or power law potential with interaction in the dark sector and satisfying $j=1$ belong to a set of zero measure.  Therefore, we can safely rule out interacting dark sector models based on non-phantom fluids and canonical scalar fields as candidates of $\Lambda$CDM-like evolution. Our present work serves to show that, although the condition $j=1$ may allow for other cosmologies kinematically degenerate with $\Lambda$CDM, the requirement for the existence of a stable future attractor dominated by non-phantom matter component may severely reduce the number of such available models. This point should always be kept in mind while trying to reconstruct alternative models based on kinematic degeneracy with $\Lambda$CDM.

Our paper is organized as follows. In Sect.\ref{ss2.5} we provide some remarks about the generic solutions of the cosmographic condition $j=1$, which is the center of our analytical treatment in this paper, and discuss why a globally spatially curved cosmology cannot attain this condition.  In Sect.\ref{ss2A} we construct a model with interactions between  dark matter and dark energy in which the energy flow is determined by imposing the cosmographic condition $j=1$; we consider both the cases of  dark energy described by a constant and a time-varying equation of state parameter and show that in both cases the null energy condition is violated. In Sect.\ref{ss2B} we maintain the previous assumptions but we describe dark energy as a canonical scalar field with either a power-law or an exponential potential and show that the Friedmann dynamics makes the system leave the physically meaningful region $0 \leq \Omega_{\rm DE} \leq 1$ except only for a zero-measure set of solutions. From the technical perspective, our results are obtained  by  performing  a dynamical system analysis.  We conclude in Sect.\ref{ss3} by discussing the astrophysical interpretation of our results arguing that the difficulty of constructing cosmological models whose background dynamics is close to the $\Lambda$CDM one (assuming General Relativity and the Copernican principle) may call for physically different solutions of the currently observed cosmological tensions.

\section{Generic solution of the condition $j=1$}
\label{ss2.5}

Our analytical treatment in this paper centers around the cosmographic condition $j=1$, which is satisfied by the spatially flat $\Lambda$CDM model. The general solution for the scale factor $a(t)$ corresponding $j=1$ is \cite[Eq.(22)]{scalefactor}
\begin{equation}\label{aABlambda}
    a(t) = \left[A e^{\lambda t} + B e^{-\lambda t}\right]^{2/3},
\end{equation}
where $A,\,B,\,\lambda$ are integration constants. However, even the explicit knowledge of the time evolution of the scale factor is not enough for drawing out conclusions about the physical interpretation of the cosmic epochs experienced by such a universe because the interpretation of the integration constants is model-dependent. For example, no information can be obtained about the values of the matter abundance parameters and nothing can be inferred about the energy conditions of the actual fluids filling the spacetime, calling for a dynamical system analysis as we will do in this paper.

A discussion regarding the global spatial curvature of the model is now in order. One may wonder whether there can be non-flat FLRW cosmology satisfying the condition $j=1$, which is typically assigned to the \emph{spatially flat} $\Lambda$CDM cosmological model. We want to mention here that $j=1$ must always correspond to a spatially flat FLRW cosmology. The reason is that integrating back from $j=1$, one gets only two terms, scaling as $(1+z)^3$ and $(1+z)^0$ respectively. On the other hand, the spatial curvature abundance for an FLRW cosmology scales as $(1+z)^2$. While dark matter and dark energy abundances appear both in the Friedmann equation, i.e. the $00$-th component of the Einstein field equations and in the continuity equations, the curvature term $\Omega_k$ appears only in the former. Thus, while the redshift scaling of dark matter and dark energy abundance change from that in the $\Lambda$CDM model when the equation of state parameter is time-dependent and/or when interaction in the dark sector is allowed, the redshift evolution of $\Omega_k$ remains unaffected. The lack of a $\sim (1+z)^2$ term in the cosmic history (\ref{HC1C2}) therefore implies spatial flatness. In other words, the two constants $\mathcal{C}_{1,2}$ cannot be anyhow related to the spatial curvature abundance $\Omega_{k0}$ as long as $j=1$ \cite{intj2}. A spatial curvature term is of course possible if, instead of $j=1$, one takes a different parametrization for $j(z)$.  For example, integrating back the parametrization 
\begin{equation}
\label{jforcurve}
    j(z) = 1 + j_{1}\frac{(1+z)^2}{h^{2}(z)},
\end{equation}
 one arrives at (see model III of Ref.\cite{intj3}) 
\begin{equation}
    h^{2}(z) = \mathcal{C}_{1}(1+z)^3 + \mathcal{C}_2 - j_{1}(1+z)^2.
\end{equation}
As usual, ${\mathcal C}_1 = {\mathcal C}_1 (\Omega_{\rm DE0},\Omega_{m0})$ and ${\mathcal C}_2 = {\mathcal C}_2 (\Omega_{\rm DE0}, \Omega_{m0})$, while the third term \emph{may} be interpreted as the spatial curvature. The estimates of the value of the latter parameter $j_1$ are consistent with a flat universe for all the combinations of data (Supernova
Distance Modulus data, Observational Hubble Data, Baryon Acoustic Oscillation
data and CMB shift parameter data) considered without the need of imposing at priori a $\Lambda$CDM epoch at the present time \cite[Table IV]{intj3}.  The identification $\{\mathcal{C}_1,\mathcal{C}_2,j_1\}=\{\Omega_{m0},\Omega_{\Lambda0},\Omega_{k0}\}$ gives back the spatially non-flat $\Lambda$CDM model. We would like to mention that while a deviation from $j=1$ following the parametrization \eqref{jforcurve} is a necessary condition for the existence of a global spatial curvature, however, it is \emph{not} a sufficient condition. One could indeed try to reconstruct specific equations of state and interaction terms, albeit very speculative, delivering a $\sim (1+z)^2$ term in the Friedmann equation. Again, the reason is that the physical interpretation of the mathematical integration constants is model-dependent.

Before moving on to further analysis, we would once again like to make it clear what we mean by kinematic degeneracy with the spatially flat $\Lambda$CDM model, or having a $\Lambda$CDM-like evolution. We imply by a $\Lambda$CDM-like evolution the generic solution of the differential $j=1$, as given by the two parameter family of solutions in Eq.\eqref{HC1C2} or the three parameter family of solutions in Eq.\eqref{aABlambda}. The reason is that such solutions have a clearly defined  CDM limit in the asymptotic past and a clearly defined  $\Lambda$-limit in the asymptotic future.  This is apparent from Eq.\eqref{HC1C2}:  as $z\rightarrow\infty$, $H^{2}\sim(1+z)^3$, while as $z\rightarrow-1$, $H^{2}=constant$.   The condition $j=1$, of course, does \emph{not} uniquely specify the $\Lambda$CDM model itself, as the lesson from the statefinder diagnostic \cite{state} tells us. In fact, if one takes \cite[Eq.3]{state} and try to reconstruct the model based on\footnote{The statefinder parameter $r$ in \cite{state} is precisely the jerk parameter $j$.}  $j=1$, one gets the following equation for a generic dark energy equation of state $w_{\rm DE}(N)$ ($N\equiv \ln a$)
\begin{equation}\label{model}
    \frac{dw_{\rm DE}}{dN} = 3w_{\rm DE}(1 + w_{\rm DE}).
\end{equation}
One of the solutions of the above equation is of course $w_{\rm DE}=-1$ i.e. the $\Lambda$CDM model, but it is clear that there are more generic dynamical dark energy models $w_{\rm DE}=w_{\rm DE}(N)$ that satisfies the above equation. This is exactly why the authors in \cite{state} had to introduce a second statefinder parameter. Nonetheless, whatever the model is, it must definitely give rise to a cosmic evolution of the form Eqs.\eqref{HC1C2} or \eqref{aABlambda}, i.e. has clearly defined effective CDM and $\Lambda$ limits. This is because Eqs.\eqref{HC1C2} or \eqref{aABlambda} is obtained by plainly integrating backwards from the cosmographic condition $j=1$, without any reference whatsoever to any particular model that can be reconstructed from Eq.\eqref{model}. In what follows, whenever we talk about $\Lambda$CDM-like evolution or kinematic/cosmographic degeneracy with the $\Lambda$CDM model, what we inherently refer to is a cosmic evolution of the form of Eq.\eqref{HC1C2} or Eq.\eqref{aABlambda}.

\section{Phase space analysis for an FLRW universe with $j=1$}
\label{ss2}

In this section we carry out the generic phase space analysis for interacting dark matter-dark energy models which reproduce the kinematics of a spatially flat  cosmology satisfying the jerk parameter $j\equiv\frac{\dddot{a}}{aH^3}=1$. The cold dark matter is modelled as a dust fluid. We consider two cases: one where the dark energy is modelled as a (not necessarily ideal) fluid and another where it is modelled as a quintessence scalar field.

\subsection{Coupled fluid model}
\label{ss2A}

The governing equations of the system are
\begin{subequations}
\begin{eqnarray}
&& \text{Friedmann constraint:} \qquad 3H^2 = \rho_m + \rho_{\rm DE},\\
&& \text{Raychaudhuri equation:} \quad 2\dot{H} + 3H^2 = - w_{\rm DE}\rho_{\rm DE},\\
&& \text{Continuity equation:} \quad \dot{\rho}_m + 3H\rho_m = - Q,\\
&& \text{Continuity equation:} \quad \dot{\rho}_{\rm DE} + 3H(1+w_{\rm DE})\rho_{\rm DE} = Q.
\end{eqnarray}
\end{subequations}
We define the dynamical variables
\begin{equation}
    \Omega_m \equiv \frac{\rho_m}{3H^2}, \qquad \Omega \equiv \frac{\rho_{\rm DE}}{3H^2},
\end{equation}
which are related by the algebraic constraint
\begin{equation}
    \Omega_m + \Omega = 1.
\end{equation}
Let us begin by discussing the case  $w_{\rm DE}=const.$.    The resulting dynamical system is 1-dimensional and given by the equation
\begin{equation}\label{dyneq_Omegam}
    \Omega_m' = 3w_{\rm DE}\Omega_{m}(1-\Omega_m) - \frac{Q}{3H^3},
\end{equation}
where a $'$ represents derivative with respect to the logarithmic time $N\equiv \ln a$. From the Raychaudhuri equation one can calculate the deceleration parameter
\begin{equation}
\label{q1}
    q \equiv -\frac{\ddot{a}}{aH^2} = -1-\frac{\dot{H}}{H^2} = \frac{1}{2}\left[1+3w_{\rm DE}(1-\Omega_m)\right],
\end{equation}
and from the deceleration parameter one can calculate the jerk parameter as
\begin{equation}
\label{j1}
    j \equiv \frac{\dddot{a}}{aH^3} = 2q^{2}+q-q' = 1 + \frac{9}{2}w_{\rm DE}(1+w_{\rm DE})(1-\Omega_m) - \frac{w_{\rm DE}Q}{2H^3}.
\end{equation}
The same relationship between jerk parameter $j$ and  interaction term $Q$ can be derived also using literature results from the {\it reconstruction method}. We take either \cite[Eq.(4)]{recoQ1} or \cite[Eq.(16)]{recoQ2} assuming $w_{\rm DE}=const.$ and write (note that in those papers the prime  stands for a derivative with respect to the redshift $z$):
\beq
\label{caiQ}
-w_{\rm DE} Q= 2 \left[ H \left( \frac{dH}{dz}\right)^2 +H^2 \frac{d^2 H}{ dz^2}  \right] (1+z)^2 -  2(5 +3w_{\rm DE} ) (1+z) H^2 \frac{dH}{dz}  +9(1+w_{\rm DE})H^3\,.
\eeq
In fact, using $\frac{d}{dt}=-(1+z)H\frac{d}{dz}$ and $\frac{d^2}{d t^2} = H (1+z)^2 \left[ H \frac{d^2}{dz^2} + \frac{dH}{dz} \frac{d}{dz}\right] +(1+z) H^2 \frac{d}{dz}$, the previous condition can be recast as
\beq
-\frac{w_{\rm DE} Q}{H^3}= 2 \frac{\ddot H}{H^3} +6 (2 +w_{\rm DE}) \frac{\dot H}{H^2} + 9(1+ w_{\rm DE})=2j -2 -9(1+w_{\rm DE})w_{\rm DE} (1-\Omega_m)\,,
\eeq
which reproduces (\ref{j1}), and where we have used $j \equiv 1+3 \frac{\dot H}{H^2}+\frac{\ddot H}{H^3}$ and (\ref{q1}) in the last step.

Imposing the condition $j=1$ now fixes the unknown coupling function $Q$
\begin{equation}
\label{QWDE}
    \frac{Q}{3H^3} = 3(1+w_{\rm DE})(1-\Omega_m).
\end{equation}
Since $1-\Omega_m \geq 0$, in this model the direction of the energy flow is fixed during the cosmic evolution. If we take the dark energy fluid to be a non-phantom fluid ($1+w_{\rm DE}>0$), then the energy flow is always from dark matter to dark energy. If we instead take the dark energy fluid to be a phantom fluid ($1+w_{\rm DE}<0$), then the energy flow is always from dark energy to dark matter. We need to recall that the direction of the energy flow is still a topic of debate  because according to thermodynamical arguments relying on the Le Ch\^atelier-Braun principle it should occur from dark energy to dark matter \cite{braun} (with this proposal being consistent with the non-parametric reconstruction of the interaction term from astrophysical datasets \cite{brauncc}), while others suggest that its direction should be opposite because the dark matter dominated epoch (in which astronomical structures form) is expected to come before the dark energy dominated one \cite{salvatelli}. 

We remark that the kinematic degeneracy at the background level with $\Lambda$CDM should be imposed by requiring either $j(z)\equiv 1$ or the cosmic history (\ref{HC1C2}) with no prejudicial physical interpretation of those constants. Actually, implementing the latter into (\ref{caiQ}), one obtains that  
\begin{equation}
    {\mathcal C}_2= - \frac{w_{\rm DE}Q}{9 H H_0^2 ( 1+w_{\rm DE})} = 1- {\mathcal C}_1,
\end{equation}
i.e. $Q(z) \propto H(z)$, within our model. Using (\ref{QWDE}) and specifying all the  quantities at the present time $z=0$, we can identify
\begin{equation}\label{generic_C}
    \lbrace {\mathcal C}_1 , {\mathcal C}_2 \rbrace = \lbrace 1 + w_{\rm DE} \Omega_{0} , - w_{\rm DE} \Omega_{0} \rbrace \,.
\end{equation}
We should now remark that, as found in Sect.\ref{ssintro}, for the actual $\Lambda$CDM model we need to have  $\{\mathcal{C}_1,\mathcal{C}_2\}=\{\Omega_{m0},\Omega_0\}$. Implementing this requirement into Eq.(\ref{generic_C}) and solving for $w_{\rm DE}$, we consistently obtain $w_{\rm DE}=-1$. Therefore, a model with $w_{\rm DE}=const. \neq -1$ and $Q \neq 0$ is {kinematically but not dynamically} equivalent to $\Lambda$CDM.

Implementing the coupling function obtained from Eq.(\ref{QWDE}) into the dynamical equation (\ref{dyneq_Omegam}) gives
\begin{equation}
\label{1dimev}
    \Omega_m' = -3(1-\Omega_m)\left[1+w_{\rm DE}(1-\Omega_m)\right].
\end{equation}
The previous equation is equivalent to
\beq
\label{1dimevb}
\Omega ' = 3\Omega (1+w_{\rm DE} \Omega)
\eeq
and it can be analytically solved as
\beq
\label{solfluid}
\Omega= \frac{1}{ {\mathcal C} e^{-3N} -w_{\rm DE} } \qquad \Rightarrow \qquad \Omega_m=\frac{{\mathcal C} e^{-3N} -w_{\rm DE} -1} {{\mathcal C} e^{-3N} -w_{\rm DE}}
\eeq
where $\mathcal C = \frac{1}{\Omega_0}+w_{\rm DE}$ is a constant of integration, and a subscript $0$ indicates that the quantity is evaluated at the present time. Two equilibrium epochs can be identified during the cosmic evolution 
\begin{equation}
\label{eqideal}
    \mathcal{P}_1:(\Omega_m,\Omega)=(1,0), \qquad \mathcal{P}_2:(\Omega_m,\Omega)=\left(\frac{1+w_{\rm DE}}{w_{\rm DE}},-\frac{1}{w_{\rm DE}}\right),
\end{equation}
of which the first one is clearly a CDM dominated epoch. Notice that, since $0<\Omega_m,\Omega<1$ the second fixed point, $\mathcal{P}_2$ can be physical iff $1+w_{\rm DE}<0$, i.e. when the dark energy fluid is phantom. It can be relevant as a solution of the coincidence problem if we choose $w_{\rm DE} \sim -2$, because in that case the abundance of dark energy and of dark matter would be the same \cite{coincidence} (more precisely, for reproducing the commonly accepted value of dark energy abundance $\Omega=0.70$, we should require $w_{\rm DE}\approx -1.43)$. Furthermore, we can introduce the equation of state parameter for the {\it effective} fluid driving the expansion of the universe, which is defined as
\beq
\label{weff}
w_{\rm eff}:=\frac{p_{\rm DE}}{\rho_m +\rho_{\rm DE}}\,.
\eeq
In correspondence of $\mathcal{P}_2$ we get $w_{\rm eff}=-1$ implying a de Sitter evolution. For establishing the stability of these two equilibria we begin by computing the Jacobian for Eq.(\ref{1dimevb}), getting $J=3(1+2 w_{\rm DE} \Omega)$. Then, we have
\beq
J\bigg\vert_{\mathcal{P}_1}=3>0 \,, \qquad J\bigg\vert_{\mathcal{P}_2}=-3<0\,.
\eeq
Therefore, the CDM dominated epoch $\mathcal{P}_1$ is a past attractor whereas the other equilibrium, $\mathcal{P}_2$, is a future attractor.
It should be noted that viable coupled fluid-fluid models satisfying $j=1$ needs to be based on  a phantom fluid with $w_{\rm DE}<-1$ for having a future attractor, without which the dynamics (\ref{solfluid}) would leave the physical acceptable region $0 \leq \Omega \leq 1$ in the future. We recall that  in the $\Lambda$CDM model there is a past attractor accounting for the matter dominated era with $\Omega_m=1$ and a future attractor given by the de Sitter universe with $\Omega_m=0$: their   existence and stability properties can be easily obtained applying our previous analysis to (\ref{1dimevb}) with $w_{\rm DE}=-1$. Therefore, the property of $\Lambda$CDM   of  exhibiting a future attractor, which is indeed necessary for having a physically meaningful dynamics, is {\it not general } and it can be used for ruling out already at the background level some models kinematically degenerate with it.

We mention that, by the very definition of the equilibrium points, it is mathematically possible to have cosmological solutions that just reside  at the equilibrium points $\mathcal{P}_1$ or $\mathcal{P}_2$. Such solutions correspond to $q(z)=\frac{1}{2}$ and $q(z)=-1$ respectively. Even though they satisfy $j(z)=1$ identically, these two specific solutions do not actually follow a $\Lambda$CDM-like evolution. The solution residing at $\mathcal{P}_1$ is an SCDM scenario ($\{r,s\}=\{1,1\}$ in the statefinder diagram of \cite{state}).

We have demonstrated that a model kinematically degenerate with $\Lambda$CDM, if based on an ideal fluid description of dark energy, e.g. with pressure and energy density directly proportional to each other,  would violate the null energy condition. We will now show  that this picture does not change even if we allow for a time evolution of the dark energy equation of state parameter. In fact, we get
\begin{equation}
j = 1 + \frac{9}{2}w_{\rm DE}(1+w_{\rm DE})(1-\Omega_m) - \frac{3}{2}(1-\Omega_m) w_{\rm DE}' - \frac{w_{\rm DE}Q}{2H^3}\,,
\end{equation}
and therefore
\beq
\label{Qnoideal}
\frac{Q}{3H^3} = \left[ 3(1+w_{\rm DE}) - \frac{w_{\rm DE}'}{w_{\rm DE}} \right](1-\Omega_m) \,,
\eeq
should we impose $j = 1$.  Under these assumptions, the evolution of dark energy abundance is governed by
\beq
\label{omeganoid}
\Omega' = \Omega\left[3(1+w_{\rm DE} \Omega) - \frac{w_{\rm DE}'}{w_{\rm DE}}\right]\,.
\eeq
A time-dependent equation of state parameter can be interpreted as being energy dependent. In other words, what we are now considering is that dark energy is a non-ideal fluid whose equation of state is $p_{\rm DE}=w(\rho_{\rm DE})\rho_{\rm DE}$. The dynamical system we should now tackle would be constituted by (\ref{omeganoid}) and by a specific differential equation for $w_{\rm DE}'$; the latter would become known if a specific modeling of dark energy is assumed from the outside. However, regardless further assumptions on the dark energy equation of state beyond the property that it is time-dependent, we know that an equilibrium point would necessarily be characterized by the property $(w_{\rm DE}')_{\rm eq}=0$. Therefore, from the evolution of dark energy abundance we can identify the equilibria,
\begin{equation}
    \mathcal{P}_1:(\Omega_m,\Omega)=(1,0), \qquad \mathcal{P}_2:(\Omega_m,\Omega)=\left(\frac{1+w_{\rm DE *}}{w_{\rm DE * }},-\frac{1}{w_{\rm DE *}}\right),
\end{equation}
where $w_{\rm DE * }$ is a constant whose explicit value would be known from solving $(w_{\rm DE}')_{\rm eq}=0$. In this scenario, $\mathcal{P}_2$ might actually represent more than one equilibrium point depending on the number of different roots of the evolution equation for the dark energy equation of state.  Even without knowing those solutions explicitly, we can conclude that the latter equilibrium point can exist only if dark energy, at $\mathcal{P}_2$, comes in the form of a phantom fluid, e.g. with $w_{\rm DE *}<-1$, consequently violating the null energy condition. A plethora of time-varying dark energy equations of state $p_{\rm DE}=w(\rho_{\rm DE})\rho_{\rm DE}$ have been investigated in literature \cite{fluid1,fluid2,fluid3,fluid4,fluid5}, with Chaplygin gas being perhaps the most popular one \cite{cg,gcg}; even allowing for extra degrees of freedom in terms of energy flows, our analysis reveals that they might be a solution of some cosmological tensions  only at the price of violating the null energy condition. Our result applies  to an interaction term (\ref{Qnoideal}) whose physical properties have been left at a quite general level. 

A curvature term inside the Friedmann constraint will change the available range for $\Omega$ potentially allowing to fulfill the null energy condition. However, as we explained in Sect.\ref{ss2.5},  the condition $j=1$ on which our models are based is consistent with spatial flatness only. Hence, a curvature term cannot be invoked for resolving this energy condition violation.

\subsection{Coupled quintessence model}
\label{ss2B}

We will now consider the scenario of dark energy being accounted for by a  quintessential scalar field $\phi$ with potential $V=V(\phi)$ and kinetic energy $K=\frac{\dot \phi^2}{2}$. The energy density and pressure of dark energy are:
\beq
\label{rhocanonical}
\rho_{\rm DE}= \frac{\dot \phi^2}{2} +V(\phi)  \,, \qquad p_{\rm DE}=\frac{\dot \phi^2}{2}-V(\phi)\,.
\eeq
The Friedmann equation and the continuity equations are now given by
\begin{subequations}\label{ffq}
\begin{eqnarray}
&& \text{Friedmann constraint:} \quad 3H^2 = \rho_m + \frac{\dot \phi^2}{2} +V(\phi), \label{ffq_fried}\\
&& \text{Raychaudhuri equation:} \quad \dot H = -\frac{\rho_m + \dot \phi^2}{2},\\
&& \text{Continuity equation:} \quad \dot \rho_m +3H  \rho_m=-Q,\\
&& \text{Continuity equation:} \quad \dot \phi \ddot \phi +V_{,\phi} \dot \phi +3H \dot \phi^2 =Q.
\end{eqnarray}
\end{subequations}

Let us now introduce the dimensionless variables \cite[Eqs.(4.14)-(4.15)]{tamanini}
\beq
\label{defxq}
x := \frac{\dot \phi}{\sqrt{6}H}\,, \qquad y:=\frac{\sqrt{|V(\phi)|}}{\sqrt{3}H}\,, \qquad \Omega_\phi:= \frac{\rho_{\rm DE}}{3H^2}= x^2 +y^2   \,, \qquad \Omega_m := \frac{\rho_m}{3H^2}\,,
\eeq
which have been proposed in \cite{copeland}, and the auxiliary quantities \cite[Eqs.(4.26),(4.28)]{tamanini}
\beq
\label{deflambda}
\lambda := -\frac{V_{, \phi}}{V}\,, \qquad \Gamma :=\frac{V V_{, \phi\phi}}{V^2_{, \phi}}\,.
\eeq
We can note that the case $\Gamma=1$ can be integrated to give the exponential potential $V(\phi)=V_0 e^{-\lambda \phi}$, where $\lambda$ and $V_0$ are constants, while for the case $\Gamma={\mathcal A}\neq1$ we would get $V(\phi)= \left(\frac{1}{(1-{\mathcal A})({\mathcal A}_1 \phi + {\mathcal A}_2)} \right)^{\frac{1}{{\mathcal A}-1}}$ with ${\mathcal A}_1$ and ${\mathcal A}_2$ being constants of integration. In the latter case, for $\phi \to 0$ the potential approaches a constant, while for $\phi \to \infty$ the potential can either diverge (if ${\mathcal A}<1$) or asymptotically vanish (if ${\mathcal A}>1$) with the value ${\mathcal A}=1$ constituting a bifurcation. Scalar field potentials exhibiting these characteristics have been playing an important role in literature, with the exponential potentials being considered because of their relations with super-gravity theories \cite{copeland,potentials1,potentials2,potentials3,potentials5}, power-law potentials exhibiting bifurcations in their asymptotic behavior depending on some free parameters arise in the context of the so-called \lq\lq pole dark energy" models \cite{potentials6},  and finally the study of  inverse power-law potentials has been pioneered by Peebles \& Ratra \cite{potentials7}.    

The dark energy (or equivalently the quintessence) equation of state parameter can be recast as \cite[Eq.(4.18)]{tamanini}
\beq
w_{\rm DE}\equiv w_\phi= \frac{x^2 -y^2}{x^2 +y^2}\,,
\eeq
while the {\it effective} equation of state parameter (\ref{weff}) now reads as
\beq
\label{weffq}
w_{\rm eff}=  x^2 -y^2 \,,
\eeq
where we have used the Friedmann  constraint given below.

The Friedmann equation (\ref{ffq_fried}) now gives the constraint
\beq
\label{friedq}
\Omega_m + \Omega_\phi = \Omega_m + x^2 + y^2 = 1\,,
\eeq
while the evolution equations are given by 
\begin{subequations}
\begin{eqnarray}
\label{xprime1}
x' &=& \frac{3x (x^2 -y^2 -1) +\sqrt{6}\lambda y^2}{2} +\frac{Q}{6H^3 x}\,, \\
y' &=&   \frac{y[3(x^2 -y^2 +1) -\sqrt{6}\lambda x]}{2}  \,,\\
\lambda ' &=& -\sqrt{6} x \lambda^2 (\Gamma -1)    \,.
\end{eqnarray}
\end{subequations}
Here we should note that, since $0 \leq \Omega_m \leq 1$, we have $0 \leq \Omega_\phi \equiv x^2 + y^2 \leq 1$ and the variables $x,y$ are already bounded: $-1 \leq x,y \leq 1$.

It should be appreciated that the interaction term affects the $x$-equation only. The third equation admits the equilibria $x=0$ which corresponds to the case of a cosmological constant $p_{\rm DE}=-\rho_{\rm DE}$, $\lambda=0$ which corresponds to the case of a constant potential $V(\phi)=V_0$, and $\Gamma=1$ which corresponds to the case of the exponential potential (for which, we  recall $\lambda=const.$). The deceleration parameter reads as
\beq
q= \frac{1+ 3(x^2 -y^2)}{2} =\frac{1+ 3 w_{\rm eff}}{2}   \,.
\eeq
Thus
\beq
q' = \frac{9 [(x^2 -y^2)^2 - (x^2+y^2)]}{2}+3\sqrt{6}\lambda x y^2 +\frac{Q}{2H^3}
\eeq
which implies
\beq
j= 1+9x^2 - 3 \sqrt{6} \lambda x y^2- \frac{Q}{2H^3}      \,.
\eeq
The condition of kinematic degeneracy with $\Lambda$CDM model $j=1$ provides
\beq
\label{Qphi}
\frac{Q}{2 H^3}= 3x(3x -  \sqrt{6} \lambda  y^2)  \,,
\eeq
where we can note that for the case of a cosmological constant for which $x=0$, we correctly reduce to the non-interacting scenario. We have $Q>0$ if either the two conditions $x>0,\,x>\sqrt{6}\lambda y^2/3$ are simultaneously met, or the two conditions $x<0,\,x<\sqrt{6}\lambda y^2/3$ are simultaneously met. The flow of energy is changing direction at $x=0$ and at $x=\sqrt{6}\lambda y^2/3$.

By using (\ref{rhocanonical}),(\ref{defxq}) we can re-write the interaction term as
\beq
Q = 3H(\rho_{\rm DE} + p_{\rm DE}) +\lambda(p_{\rm DE} -\rho_{\rm DE})\sqrt{\rho_{\rm DE} + p_{\rm DE}}
\eeq
which is shown to be non-linear in the dark energy density. It has been argued in \cite{chimento} that  non-linear energy exchanges in the dark sector give rise to a cosmological dynamics similar to that of a universe filled with a Chaplygin Gas: we confirm in Table \ref{table_eqpts1} that this is the case even when $\lambda$ is not anymore a fixed parameter but a dynamical variable by identifying the past dark matter dominated epoch and the future de Sitter attractor.

Implementing (\ref{Qphi}) into (\ref{xprime1}) we obtain
\beq
x'=   \frac{3x (x^2 -y^2 +1)-\sqrt{6} \lambda y^2}{2}\,.
\eeq
The dynamical system to consider therefore is
\begin{subequations}\label{dsa_quint}
\begin{eqnarray}
\label{xprime2}
x' &=& \frac{3x (x^2 -y^2 +1)-\sqrt{6} \lambda y^2}{2}\,, \\
\label{yprime}
y' &=&   \frac{y[3(x^2 -y^2 +1) -\sqrt{6}\lambda x]}{2}  \,,\\
\label{lambdaprime}
\lambda ' &=& -\sqrt{6} x \lambda^2 (\Gamma -1)    \,.
\end{eqnarray}
\end{subequations}
Once a functional form of the scalar field potential is given, $\Gamma$ can be expressed as a function of $\lambda$, provided that the definition $\lambda=-\frac{V_{, \phi}}{V}$ is invertible. In particular, the situation is simplified when $\Gamma$ is a constant, and even further if $\Gamma=1$ (the case of exponential potential), for which the phase space is reduced to 2-dimensions. We note that the dynamical system \eqref{dsa_quint} is always symmetric under the transformation $(y\mapsto-y)$, so that we can focus our attention only to the region $0\leq y\leq1$, i.e. where the potential is nonnegative. For the special cases when $\Gamma$ is either constant or an even function of $\lambda$, the dynamical system \eqref{dsa_quint} is also symmetric under the simultaneous transformation $(x,\lambda)\mapsto(-x,-\lambda)$, so that we can constrain our region of focus even further to  $\lambda>0$. These symmetries characterize as well the non-interacting case of the coupled quintessence models \cite[Sects. 4.3-4.4]{tamanini}, although the numbers and stability properties of equilibria differ among the two scenarios. We must mention that the symmetries of the non-interacting scenarios are in general not preserved when an interaction term is added. Since the term $\frac{Q}{6H^{3}x}$  can be in fact any arbitrary function of the dynamical variables $x,\,y,\,\lambda$ and there is no guarantee that the symmetries $y\mapsto - y$ and $(x,\lambda)\mapsto(-x,-\lambda)$ will be preserved even when an arbitrary interaction term is involved. It is, in fact, remarkable that the particular interaction term specified by the requirement of kinematic degeneracy with $\Lambda$CDM, namely the one given by Eq.(\ref{Qphi}), does in fact preserve the symmetry of the non-interacting case.

\begin{figure}
	\begin{center}
		$
		\begin{array}{ccc}
		{\includegraphics[angle=0,scale=0.45]{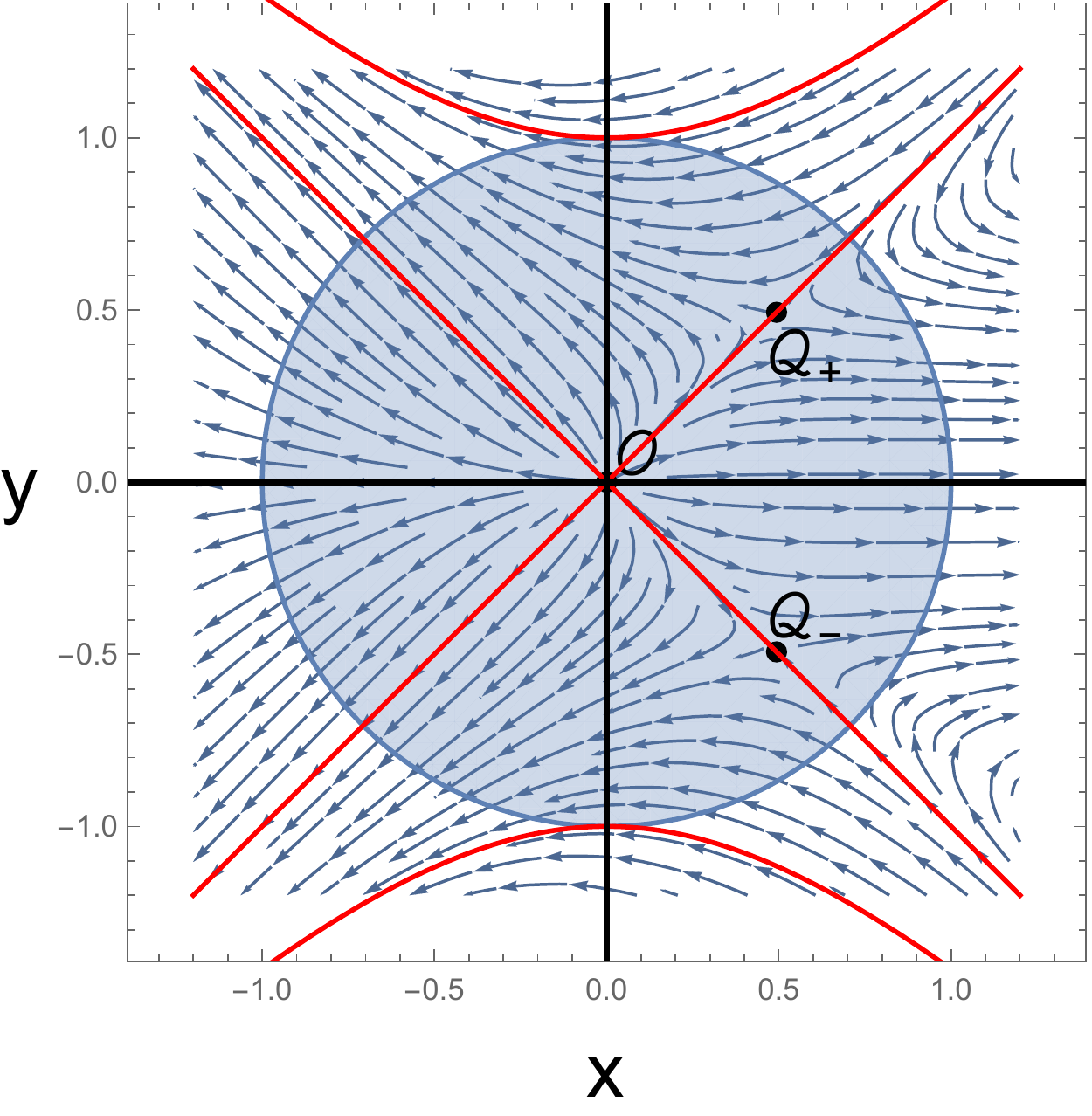}}  &
		{\includegraphics[angle=0,scale=0.45]{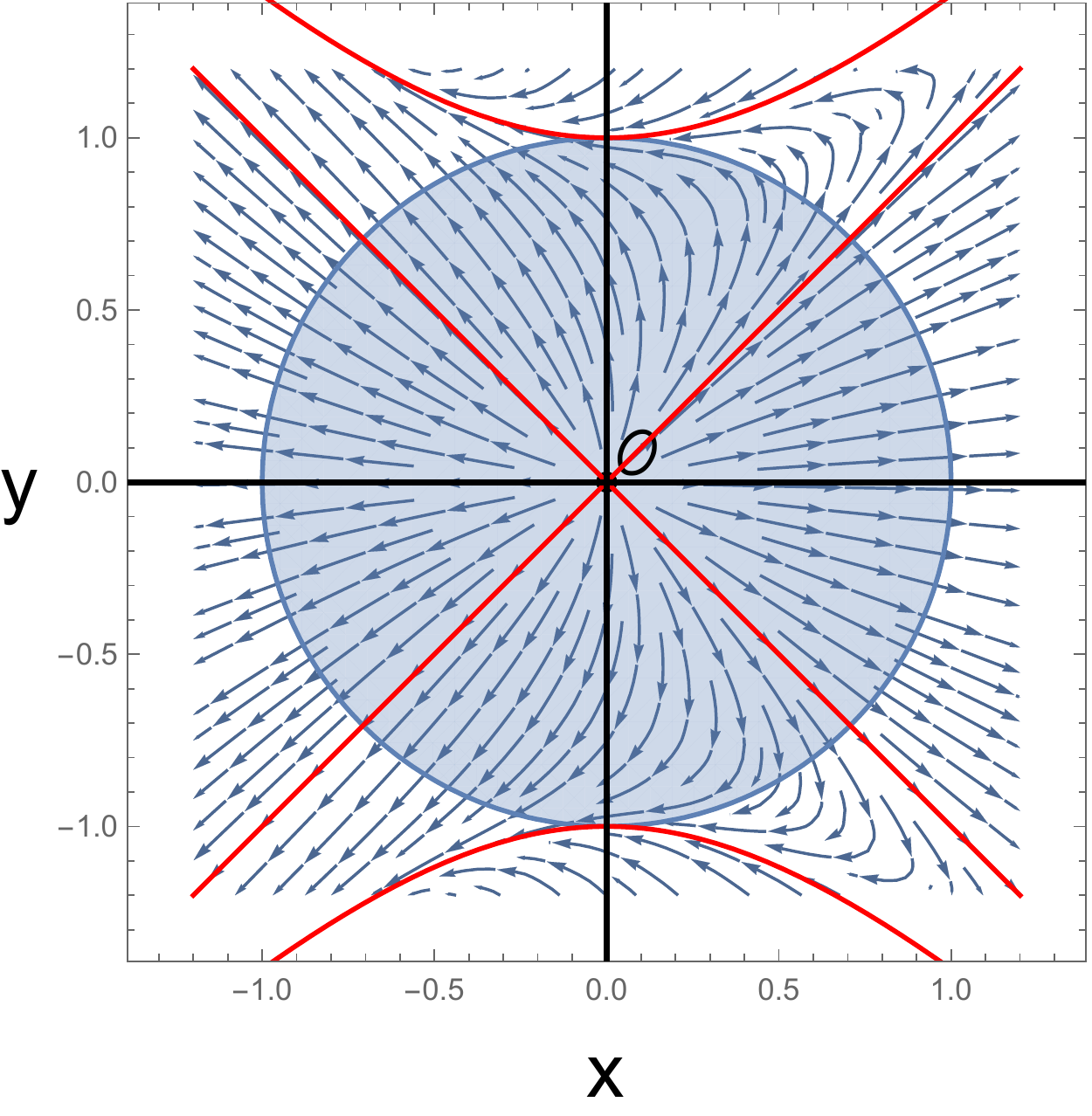}}  &
		{\includegraphics[angle=0, scale=0.45]{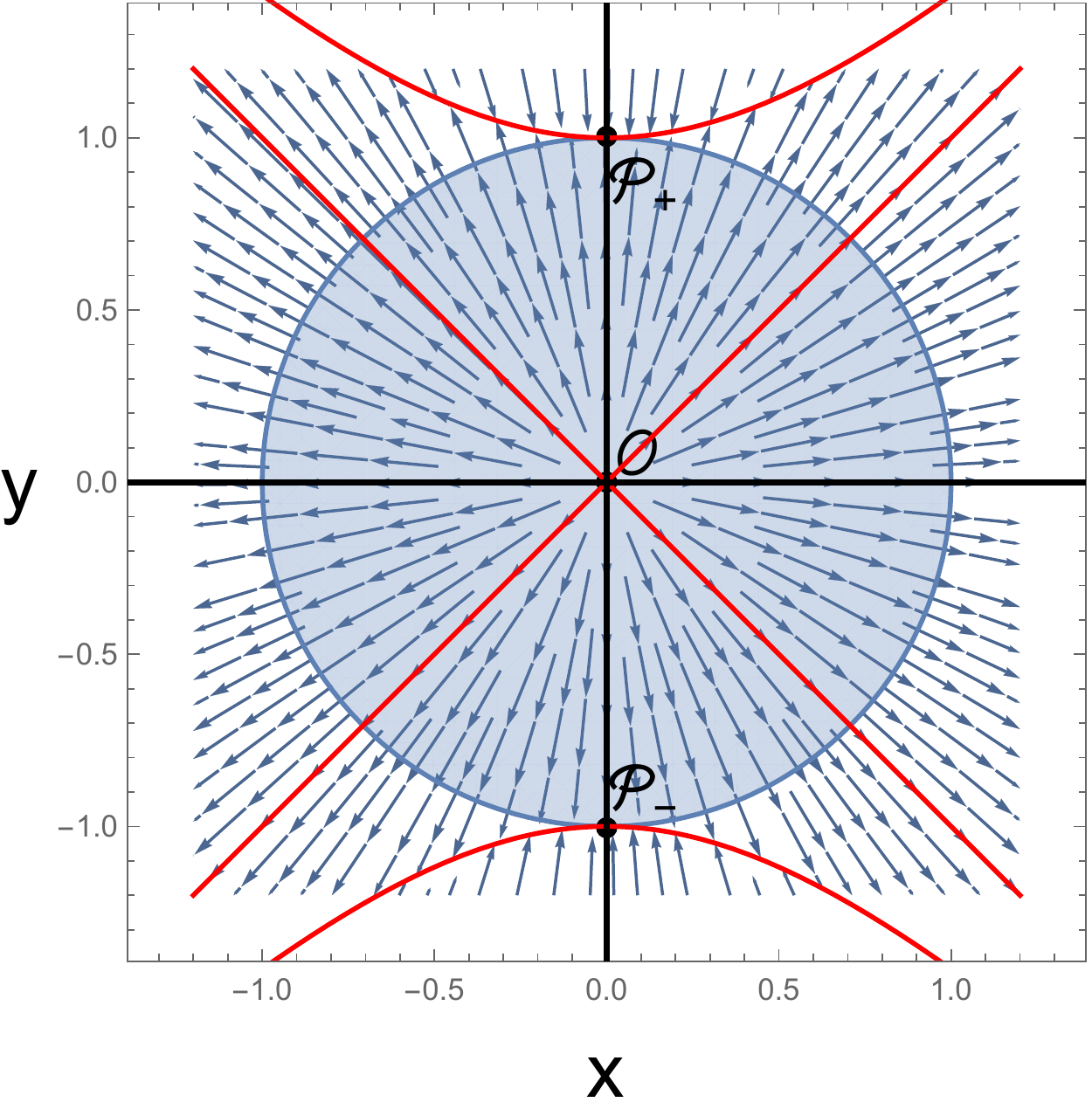}} \\ 
		(a) & (b) & (c)
		\end{array}$
	\end{center}
    \caption{In panel (a),(b),(c) we show the 2-D phase space dynamics created by the dynamical system (\ref{xprime2}) and (\ref{yprime}), fixing $\lambda=2.5$, $\lambda=0.5$ and $\lambda=0$, respectively. The invariant submanifolds $s(x,y)=0,1$ where $s(x,y)\equiv x^{2}-y^{2}+1$ are represented by red lines and curves. These invariant submanifolds, along with the invariant submanifold $y=0$, split the full phase space into disjoint regions in which the dynamics remains confined depending on the initial condition. For $\lambda=0$, the hyperbola $s(x,y)=0$ also constitutes a 1-parameter family of non-isolated fixed points. The shaded unit circle corresponds to the region of physical viability as given by $0\leq x^{2}+y^{2}=1-\Omega_m\leq1$. These phase portraits show the actual 2-D phase space dynamics for exponential potential (the case $\Gamma=1$) and 2-D projections on various $\lambda$-planes of the 3-D phase space dynamics for power law potential (the case $\Gamma=\mathcal{A}\neq1$). The plots confirm  the results exhibited in Table \ref{table_eqpts2} and in the main text: {\it (i)} the equilibrium point ${\mathcal O}$ exists regardless the chosen value for $\lambda$, unlike ${\mathcal P}_{\pm}$ and ${\mathcal Q}_{\pm}$; {(\it ii)} ${\mathcal O}$ is a source, ${\mathcal P}_{\pm}$ are local attractors and ${\mathcal Q}_{\pm}$ are saddles. We mention that the points $\mathcal{P}_{\pm}$ exist for both power law and exponential potential (see both Table \ref{table_eqpts1} and Table \ref{table_eqpts2}), and the point $\mathcal{O}$ resides on the line of fixed points $\mathcal{L}$ of Table \ref{table_eqpts1}. Although the physically viable region is the one inside the shaded circle, we also show the region outside it to confirm that the coupled quintessence cosmologies kinematically degenerate to $\Lambda$CDM do in fact always go outside the shaded region, rendering them physically non-viable. There are two very special fine tuned cases for which there can be physically viable coupled quintessence cosmologies kinematically degenerate to $\Lambda$CDM: {\it (i)} Exponential potential with $\lambda=0$ (i.e. a constant potential) and {\it (ii)} Power law potential with the $\lambda=0$ invariant submanifold being attracting. Both these cases actually reduce to a canonical scalar field cosmology with constant potential, which is represented by the two vertical lines connecting the points $\mathcal{O}$ to $\mathcal{P}_{\pm}$ in panel (c).}
    \label{fig2d}
\end{figure}

For $\Gamma=\mathcal{A}\neq1$, the dynamical equations \eqref{xprime2},\eqref{yprime},\eqref{lambdaprime} create a 3-D phase portrait. We would like to mention that, in the phase space, the specific solutions residing entirely on the invariant submanifolds $y=\pm x$ or $y^2 = 1+x^2$, do not correspond to a $\Lambda$CDM-like evolution even though they are perfectly viable mathematical solutions for the condition $j(z)=1$. Such solutions correspond to $q(z)=\frac{1}{2}$ and $q(z)=-1$ respectively. The solution residing on the invariant submanifold $y=\pm x$ is an SCDM scenario ($\{r,s\}=\{1,1\}$ in the statefinder diagram of \cite{state}).  Fig.\ref{fig2d} shows the 2-D projections of this phase portrait on three different planes corresponding to three different values of $\lambda$, revealing that the system always leaves the physically viable region except only for a zero measure set of initial conditions. 

\subsubsection{$\Gamma=\mathcal{A}\neq1$; $V(\phi)= \left(\frac{1}{(1-{\mathcal A})({\mathcal A}_1 \phi + {\mathcal A}_2)} \right)^{\frac{1}{{\mathcal A}-1}}$}

Since the case $\Gamma=const.$ already allows us to explore some cosmologically relevant scenarios, as previously mentioned, we list in Table \ref{table_eqpts1} the physically viable equilibrium points for the dynamical system formed by Eqs.(\ref{xprime2}),(\ref{yprime}),(\ref{lambdaprime}), treating $\Gamma$ as a parameter and $\lambda$ as a dynamical variable.
\begin{table}
    \begin{center}
    	\begin{tabular}{|c|c|c|c|c|c|c|c|c|}
    		\hline
    		&&&&&&&&
    		\\
    		 Fixed point & $x_{\rm eq}$ & $y_{\rm eq}$ & $\lambda_{\rm eq}$ & $\Omega_m$ & $w_{\rm eff}$ & Cosmology & Eigenvalues & Stability
    		\\
    		&&&&&&&& \\
    		\hline 
    		&&&&&&&&
    		\\
    		 $\mathcal{L}$ & $0$ & $0$ & $\lambda_{\rm eq}$ & $1$ & $0$ & Matter Dominated power law ($a(t)\sim t^{2/3}$) & $\left( \frac{3}{2}, \, \frac{3}{2}, \, 0 \right)$ & Unstable
    		\\ 
    		&&&&&&&&
    		\\
    		 $\mathcal{P}_{\pm}$ & $0$ & $\pm1$ & $0$ & $0$ & $-1$ & Scalar field dominated de Sitter & $\left( -3, \, 0, \, 0 \right)$ & Saddle or stable
    		\\
    		&&&&&&&&
    		\\
    		\hline
            \end{tabular}
        \end{center}
    \caption{In this Table, we list the equilibrium points for the dynamical system formed by Eqs.(\ref{xprime2}),(\ref{yprime}),(\ref{lambdaprime}), where the definitions of the variables $x$, $y$ and $\lambda$ are given in Eqs.(\ref{defxq}),(\ref{deflambda}). A cosmology reaches either the point $\mathcal{P}_+$ or the point $\mathcal{P}_-$ depending on the initial conditions. The cold dark matter abundance and the effective equation of state parameter are then computed from (\ref{weffq}) and (\ref{friedq}), respectively. We recall that this scenario corresponds to have a scalar field obeying to a power-law potential.}
    	\label{table_eqpts1}
    \end{table}
Under the same assumption, the Jacobian of the system can be computed to be
\beq
J=
\begin{pmatrix}
\frac{3 (3 x^2 -y^2 +1)}{2} & -y(3x +\sqrt{6} \lambda)& -\frac{\sqrt{6} y^2}{2}\\
\frac{y(6x -\sqrt{6}\lambda)}{2} & \frac{3(x^2-3y^2+1)-\sqrt{6} \lambda x}{2} & -\frac{\sqrt{6}xy}{2} \\
\sqrt{6} \lambda^2(1-\Gamma) & 0 & 2 \sqrt{6} x \lambda (1-\Gamma)
\end{pmatrix}_{x=x_{\rm eq},\,y=y_{\rm eq},\,\lambda=\lambda_{\rm eq}} \,.
\eeq
We also list the jacobian eigenvectors corresponding to the fixed points in Table \ref{tableeigenvect}.
\begin{table}
    \begin{center}
    	\begin{tabular}{|c|c|}
    		\hline
    		&
    		\\
    		 Fixed point & Eigenvectors 
    		\\
    		& \\
    		\hline 
    		&
    		\\
    		 $\mathcal{L}$ & $
\begin{pmatrix}
\sqrt{\frac{3}{2}}\frac{1}{2(1-\Gamma) \lambda^2}\\
0 \\
1\end{pmatrix}
$  $
\begin{pmatrix}
0\\
1 \\
0\end{pmatrix}
$ $
\begin{pmatrix}
0\\
0 \\
1\end{pmatrix}
$ 
    		\\ 
    		&
    		\\
    		 $\mathcal{P}_\pm$ & $
\begin{pmatrix}
0\\
1\\
0\end{pmatrix}
$  $
\begin{pmatrix}
1\\
0 \\
0\end{pmatrix}
$ $
\begin{pmatrix}
0\\
0 \\
0\end{pmatrix}
$ 
    		\\
    		\hline
            \end{tabular}
        \end{center}
    \caption{In this Table, we list the eigenvectors corresponding to the fixed points in Tab.\ref{table_eqpts1}.}
    	\label{tableeigenvect}
\end{table}

As equilibria, we find a line of non-isolated fixed points ${\mathcal L}$  parametrized by $\lambda$  which correspond to a matter dominated universe.  Two eigenvalues are positive, while the zero eigenvalue corresponds to the eigenvector $\begin{pmatrix}
0\\
0 \\
1\end{pmatrix}$ which is aligned along the $\lambda$-direction, e.g. it is tangent to the set. Therefore, ${\mathcal L}$ is a {\it normally hyperbolic set} and it constitutes the possible past attractors of this cosmological model since all the non-zero eigenvalues are positive \cite[p.24]{coleybook}. 

Furthermore, we note that the two branches of the hyperbola
\beq
\mathcal{H}\equiv(\lambda,x^{2}-y^{2}+1)=(0,0)
\eeq
also constitute a 1-parameter family of non-isolated equilibrium points. However, taking into account the physical constraint $0\leq x^{2}+y^{2}\leq 1$, we also note that the two points $\mathcal{P}_\pm\equiv(x_{\rm eq},y_{\rm eq})=(0,\pm1)\subset\mathcal{H}$ are the only physically viable fixed points. These two equilibria are mutually exclusive, residing on the opposite sides of the invariant submanifold $y=0$. The eigenvalues of the Jacobian at $\mathcal{P}_\pm$ are $-3,0,0$. One of the zero eigenvalues corresponds to the eigenvector $\begin{pmatrix}
1\\
0 \\
0\end{pmatrix}$, which is tangent to $\mathcal{H}$ at $\mathcal{P}_\pm$. Therefore this zero eigenvalue appears because the points $\mathcal{P}_\pm$ actually belong to the 1-parameter family of nonisolated fixed points $\mathcal{H}$. The eigenvector corresponding to the other zero eigenvalue, however, comes out to be $\begin{pmatrix}
0\\
0 \\
0\end{pmatrix}$, implying that it is not even possible to straightforwardly apply the center manifold analysis. We need to devise some other technique to find out the stability of $\mathcal{P}_\pm$.

To this goal, we first note that the hyperbola $\mathcal{H}$ lies entirely on the 2-dimensional invariant submanifold $\lambda=0$. Moreover, the 2-dimensional surface given by
\beq\label{def_s}
s(x,y)\equiv x^{2}-y^{2}+1=0
\eeq
is also a 2-dimensional invariant submanifold. This is evident if we calculate
\beq\label{stab_s}
s'\equiv2(xx'-yy')=3s(s-1),
\eeq
where eqs.(\ref{xprime2})-(\ref{yprime}) have been used. The 1-parameter family of fixed points $\mathcal{H}$, therefore, lies at the intersection of two 2-dimensional invariant submanifolds, namely $\lambda=0$ and $s(x,y)=0$. It is easy to see from Eq.\eqref{stab_s} that the invariant submanifold $s(x,y)=0$ is stable, which is also apparent if we calculate
\beq
\frac{\partial}{\partial s}\left(\frac{ds}{dN}\right)\Big|_{s=0} = -3 < 0.
\eeq
This shows that the phase flow component parallel to the $x$-$y$ plane near the invariant submanifold $s(x,y)=0$ is always towards it. To be more conclusive, we need a similar analysis to investigate the stability of $\lambda=0$, but the calculation is not so straightforward for this case, as
\beq
\frac{\partial}{\partial\lambda}\left(\frac{d\lambda}{dN}\right)\Big|_{\lambda=0} = 0.
\eeq
Therefore linear analysis is inconclusive to find the directionality (towards or away from) of the $\lambda$-component of the phase flow near the invariant submanifold $\lambda=0$, and typically we need to consider higher orders partial derivatives. But even without knowing the $\lambda$-component of the flow, since we already know that $s(x,y)=0$ is a stable invariant submanifold, we can conclude that the fixed points belonging to the 1-parameter family $\mathcal{H}$ are either stable or saddle fixed points. Consequently the fixed points $\mathcal{P}_\pm$ are also stable or saddle fixed points, with $\mathcal{P}_\pm$ being definitely attracting along the $x$-$y$ plane and either attracting or repelling along the $\lambda$-direction. 

From Fig.\ref{fig2d} we note that, in general any phase trajectory originating from a point on the line $\mathcal{L}$ always leaves the physically acceptable region $0\leq x^{2}+y^{2}\leq 1$ at some point in its entire evolution history, rendering the cosmology it represents to be physically non-viable. There is only one exception provided that the points $\mathcal{P}_\pm$ are attractors. In this case, the single trajectory lying entirely on the invariant submanifold $\lambda=0$ and connecting the point $(x,y)=(0,0)$ with $\mathcal{P}_\pm$ are the only physically viable cosmologies.

\subsubsection{$\Gamma=1$; $V(\phi)=V_0 e^{-\lambda \phi}$}

On the other hand, we list in Table \ref{table_eqpts2} the equilibrium points for the dynamical system formed by Eqs.(\ref{xprime2}),(\ref{yprime}) in which $\lambda$ is treated as a fixed parameter rather than as a dynamical variable.
\begin{table}
    \begin{center}
    	\begin{tabular}{|c|c|c|c|c|c|c|c|c|}
    		\hline
    		&&&&&&&&
    		\\
    		 Fixed point & $x_{\rm eq}$ & $y_{\rm eq}$ & Existence & $\Omega_m$ & $w_{\rm eff}$ & Cosmology & Eigenvalues & Stability
    		\\
    		&&&&&&&& \\
    		\hline 
    		&&&&&&&&
    		\\
    		 $\mathcal{O}$ &  $0$ & $0$ & Always & $1$ & $0$ & Matter Dominated power law ($a(t)\sim t^{2/3}$) & $\left( \frac{3}{2}, \, \frac{3}{2} \right)$ & Unstable
    		\\
    		&&&&&&&&
    		\\
    		 $\mathcal{P}_{\pm}$ & $0$ & $\pm1$ & $\lambda =0$ & $0$ & $-1$ & Scalar field dominated de Sitter & (-3, 0) & Stable
    		\\
    		&&&&&&&&
    		\\
    		 $\mathcal{Q}_{\pm}$ & $\frac{\sqrt{6}}{2 \lambda}$ & $\pm\frac{\sqrt{6}}{2 \lambda}$ & $\lambda \leq -\sqrt{3}$ or $\lambda \geq \sqrt{3}$ & $1-\frac{3}{\lambda^2}$ & $0$ & Power law ($a(t)\sim t^{\frac{2}{3}}$) & $\left( -\frac{3}{2}, \, 3 \right)$ & Saddle
    		\\
    		\hline
            \end{tabular}
        \end{center}
    \caption{In this Table, we list the equilibrium points for the dynamical system formed by Eqs.(\ref{xprime2}),(\ref{yprime}) in which $\lambda$ is  a parameter  whose range of validity is found by demanding $ 0 \leq x^2 + y^2 \leq 1$ (or equivalently $ 0 \leq \Omega_m \leq 1$). The definition of the variables $x$ and $y$  is given in Eq.(\ref{defxq}). The cold dark matter abundance and the effective equation of state parameter are then computed from (\ref{weffq}) and (\ref{friedq}), respectively. It should be noted that in proximity of the equilibria $\mathcal{Q}_\pm$ the scale factor  is {\it effectively} evolving as in a (pressureless) dust universe (whose behavior is obtained by considering the effective equation of state parameter), although there might not be dust at all filling the space, as seen from the value of the matter parameter $\Omega_m$. We confirm graphically in Fig. \ref{fig2d} the stability results  reported here.}
    \label{table_eqpts2}
    \end{table}
We recall that this scenario corresponds to considering a scalar field governed by an exponential rather than polynomial potential. The Jacobian matrix is now
\beq
J=
\begin{pmatrix}
\frac{3 (3 x^2 -y^2 +1)}{2} & -y(3x +\sqrt{6} \lambda)\\
\frac{y(6x -\sqrt{6}\lambda)}{2} & \frac{3(x^2-3y^2+1)-\sqrt{6} \lambda x}{2} \end{pmatrix}_{x=x_{\rm eq},\,y=y_{\rm eq}} \,.
\eeq
In general for any $\lambda\neq0$, both our analytical and graphical analyses concerning the exponential potential reveal the lack of a stable fixed point inside the physically viable region $ 0 \leq \Omega_m \leq 1$, implying that the system eventually leaves this region. This result rules out the coupled quintessence model with exponential potential and effectively breaks the kinematic degeneracy with $\Lambda$CDM already at the background level. Therefore a coupled quintessence model with an exponential potential, even though exhibiting the cosmographic property $j=1$, cannot physically mimic the $\Lambda$CDM cosmology. 

Considering however the special case $\lambda=0$, just like its 3d counterpart, the two branches of the hyperbola $x^{2}-y^{2}+1=0$ correspond to a 1-parameter family of stable fixed points, which is also clear from the panel (c) of Fig.\ref{fig2d}. However, due to the physical constraint  $0 \leq x^{2}+y^{2}\leq1$ only the points ($x_{\rm eq}=0$, $y_{\rm eq} = \pm 1$) belonging to those hyperbola are physically acceptable. In this case there are only two very specific coupled quintessence cosmologies that are kinematically equivalent to the $\Lambda$CDM. In the phase portrait of panel (c), Fig.\ref{fig2d}. they correspond to the straight lines joining $\mathcal{O}$ to $\mathcal{P}_\pm$.  

\subsubsection{Discussion}

The configurations characterized by $\lambda = 0$, which can arise in both the cases of exponential and power-law potentials, essentially correspond to a constant potential, as is clear from Eq.(\ref{deflambda}).
For $\lambda=0$ one can exactly solve the dynamical equations and find the general solutions
\begin{equation}
        x(N) = \pm\frac{e^{\frac{3N}{2}}}{\sqrt{C_1^{2}e^{3N}+C_2^2}},\qquad
        y(N) = \pm\frac{\sqrt{1+C_1^2}e^{\frac{3N}{2}}}{\sqrt{C_1^{2}e^{3N}+C_2^2}},
\end{equation}
where $C_{1,2}$ are integration constants and the signs of $x(N)$ and $y(N)$ can be chosen independently of each other. As $N\rightarrow-\infty$, $(x(N),y(N))\rightarrow\left(0,0\right)$, implying that all such cosmologies have as past attractor the point $\mathcal{O}\equiv(x,y)=(0,0)$ in the phase space. As $N\rightarrow+\infty$, $(x(N),y(N))\rightarrow\left(\pm\frac{1}{|C_1|},\pm\frac{\sqrt{1+C_1^2}}{|C_1|}\right)$, implying that all such cosmologies have as the future attractor some point on the hyperbola $x^{2}-y^{2}+1=0$. However, the only cosmology that is physically viable (i.e. the straight lines joining $\mathcal{O}$ to $\mathcal{P}_\pm$) corresponds to choice $C_1\rightarrow\pm\infty$. Therefore we conclude that physically viable coupled quintessence models kinematically degenerate to $\Lambda$CDM, although in principle possible, are  extremely rare and fine tuned in the sense that there is not a family of such solutions but only two solutions, therefore constituting a set of measure zero within the available phase space.

A discussion now is in order comparing the interacting scenario we discussed here with its non-interacting counterpart \cite[Sects. 4.3,4.4]{tamanini}. Firstly, in the non-interacting case, $x^{2}+y^{2}=1$ or $\Omega_m=0$ is an invariant submanifold. Phase trajectories inside this region, which corresponds to physically viable cosmologies, never cross out of this region. This is because the dark matter density is conserved; if it is initially zero, it is zero throughout. On contrary, for the interacting case, due to the existence of an interaction term, dark matter density is not conserved; even if it is initially zero, the interaction term may induce it. All the fixed points for the non-interacting case lying on the circle $\Omega_m=0$ or $x^{2}+y^{2}=1$, apart from the points $(x,y,\lambda)=(0,\pm1,0)$, are absent in the interacting case.

\section{Conclusion}
\label{ss3}
 
Different triplets  of values of the cosmological parameters ($H_0$, $\Omega_{m0}$, $\Omega_{k0}$) can deliver the same  CMB temperature anisotropy power spectrum making estimates of the curvature of the Universe from this dataset troubled  by a geometrical degeneracy \cite{cmb1,cmb2,cmb3}. Cosmic chronometers data have therefore been exploited delivering the estimate $\Omega_k=-0.0054 \pm 0.0055$ for the curvature, if a 8-parameter model (in which also the values of  the curvature and of the sum of the neutrino masses are optimized through the statistical analysis) is assumed \cite{sunny1}. A model-independent analysis, which does not require any a priori assumption on the cosmic history of the Universe and on its matter content,  of the latter points towards a spatially flat Universe, although with a larger uncertainty delivering $\Omega_k=-0.03 \pm 0.26$ \cite{sunny2}. In the framework of General Relativity and assuming the Copernican principle, accounted for by the Friedman-Lema\^itre-Robertson-Walker metric, a spatially flat universe is consistent with the cosmographic requirement  $j=1$; conversely, this condition, which is not ruled out by astrophysical measurements according to some analyses which constrained specific redshift-dependent deviations \cite{intj1,intj2,intj3,cosmoII} \cite{reza1,reza2}, necessarily requires spatial flatness.  In this paper, we have constructed some cosmological models assuming the validity of General Relativity, of the Copernican Principle (homogeneity and isotropy at large scale) and on dark energy pictured  either by some non-ideal fluids or by canonical scalar fields interacting with dark matter. The  form of the energy coupling has been specifically re-constructed in such a way that $j=1$: this choice allows us  to reproduce a cosmic evolution at a background level consistent with that of the $\Lambda$CDM model. The need of introducing extra degrees of freedom in the forms of an evolving dark energy equation of state  and/or of an interaction arises because, assuming that data collections and analysis have been performed correctly\footnote{Although, for completeness, we need to inform about  literature studies according to which modifications to late-time Physics cannot resolve the Hubble tension because of a not appropriate use of distance ladder measurements \cite{prior}.}, some cosmological tensions such as the $f\sigma_8$ and $H_0$ tensions call for some modifications of the $\Lambda$CDM paradigm. However,  our investigation, which  exploits a dynamical system approach, has revealed the lack of physically acceptable late-time de Sitter attractors in the previously mentioned models. We astrophysically interpret our findings as suggesting that, observed cosmological tensions call for a modification of the standard $\Lambda$CDM model in some other direction, for example a modification to General Relativity or a relaxation of the Copernican principle adopting inhomogeneous cosmological modelings. 
 
A redshift-dependent rather than constant parametrization of the dark energy equation of state may alleviate the Hubble tension \cite{nature1}; when allowing also for interactions with dark matter, joint data analysis of CMB and supernovae data favors a phantom fluid over a cosmological constant \cite{diva}. More recent analysis keeps showing  that non-phantom fluids seem unable to reconcile  discrepant estimates of the Hubble constant \cite{oin}. A number of future missions are expected to narrow down the estimates of the cosmological parameters (such as the local distance ladder observations \cite{future1} and the gravitational time delays measurements \cite{future2}), and the properties of the dark energy (as the Euclid space-based survey mission \cite{future3} and the Dark Energy Spectroscopic Survey \cite{future4}).  For the time being,  in our paper we tackled in full theoretical generality the problem of constructing interacting fluid-fluid and fluid-(canonical) field models with a $\Lambda$CDM background cosmic history, allowing for extra degrees of freedom which, intuitively, should be tuned at a perturbation level for taming some observational tensions.  We arrived at the conclusion that, for a perfect fluid, the null energy condition is violated regardless the number of free parameters on which an equation of state $p_{\rm DE}=w(\rho_{\rm DE})\rho_{\rm DE}$ would be based and, for a canonical scalar field, the set of  acceptable  solutions is of measure zero.   
 
Two other interesting avenues for research allowing for such extra degrees of freedom in the theory are modifications to GR and relaxing the Copernican principle. Both these ideas have been employed as alternative to dark energy models. Along the first line of research, investigation of $f(R)$ cosmologies cosmographically equivalent to $\Lambda$CDM appears to have already been gaining attentions, bulk of which consists of attempting to reconstruct the function $f(R)$ \cite{scalefactor,f(R)1,f(R)2,f(R)3}. In Ref.\cite{f(R)5} this attempt was generalized to $f(R)$ theories non-minimally coupled to matter, which is a modified gravity model giving rise to an interaction in the dark sector. However, it was found that the reconstructed function $f(R)$ is obtained in the form of hypergeometric functions, which renders any further analytical treatment difficult. Subsequently, a dynamical system approach without the need for explicit reconstruction of the form of $f(R)$ was developed in Ref.\cite{Chakraborty}, and in our present paper we have applied the same way of thinking to a postulated energy interaction term.
Along the second line of research, Ref.\cite{ltb} investigates how far  the spherically symmetric LTB model can mimic the concordance model, discussing also the possibility of breaking the degeneracy at the perturbation level.
Therefore, in future works we will try to further deepen the understanding of the applicability of these scenarios as remedy of observed cosmological tensions by jointly applying  dynamical system techniques and  cosmographic reconstruction methods.

\begin{acknowledgments}
SC acknowledges funding support from the NSRF via the Program Management Unit for Human Resources and Institutional Development, Research and Innovation [grant number B01F650006]. D.G. is a member of the GNFM working group of Italian INDAM. D.G. acknowledges as well economical support from the start-up plan of Jiangsu University of Science and Technology.  B.M. acknowledges IUCAA, Pune, India for providing academic support through the Visiting Associateship Program.
\end{acknowledgments}


\begin{thebibliography}{99}
\section*{References}




\bibitem{intertwined1}
E.~Di Valentino, \textit{et al.},
``Snowmass2021 - Letter of interest cosmology intertwined I: Perspectives for the next decade,''
{\hypersetup{urlcolor=vividviolet}\href{https://www.sciencedirect.com/science/article/pii/S0927650521000505}{Astropart. Phys. \textbf{131} (2021) 102606}}, \href{https://arxiv.org/abs/2008.11283}{arXiv:2008.11283 [astro-ph.CO]}.

\bibitem{intertwined2}
E.~Abdalla, \textit{et al.},
``Cosmology Intertwined: A Review of the Particle Physics, Astrophysics, and Cosmology Associated with the Cosmological Tensions and Anomalies,''
{\hypersetup{urlcolor=vividviolet}\href{https://www.sciencedirect.com/science/article/pii/S2214404822000179}{J. High En. Astrophys.   \textbf{34} (2022) 49}}, \href{https://arxiv.org/abs/2203.06142}{arXiv:2203.06142  [astro-ph.CO]}.

\bibitem{intertwined22}
L.~Perivolaropoulos, F.~Skara,
``Challenges for $\Lambda$CDM: An update,"
{\hypersetup{urlcolor=vividviolet}\href{https://doi.org/10.1016/j.newar.2022.101659}{New Astronomy Reviews.   \textbf{95} (2021) 101659}}, \href{https://doi.org/10.48550/arXiv.2105.05208}
{arXiv:2105.05208  [astro-ph.CO]}



\bibitem{TT1}
\textit{Planck Collaboration},
``Planck 2018 results. VI. Cosmological parameters,''
{\hypersetup{urlcolor=vividviolet}\href{https://www.aanda.org/articles/aa/full_html/2020/09/aa33910-18/aa33910-18.html}{A\&A   \textbf{641} (2020) A6}}, \href{https://arxiv.org/abs/1807.06209}{arXiv:1807.06209  [astro-ph.CO]}; \textit{Erratum} {\hypersetup{urlcolor=vividviolet}\href{https://www.aanda.org/articles/aa/full_html/2021/08/aa33910e-18/aa33910e-18.html}{A\&A   \textbf{652} (2021) C4}}.



\bibitem{TT2}
A.~G.~Riess, S.~Casertano, W.~Yuan, J.~Bradley Bowers, L.~Macri, J.~C.~Zinn, D.~Scolnic, 
``Cosmic Distances Calibrated to 1\% Precision with Gaia EDR3 Parallaxes and Hubble Space Telescope Photometry of 75 Milky Way Cepheids Confirm Tension with LambdaCDM,''
{\hypersetup{urlcolor=vividviolet}\href{https://iopscience.iop.org/article/10.3847/2041-8213/abdbaf}{ApJ   \textbf{908} (2021) L6}}, \href{https://arxiv.org/abs/2012.08534}{arXiv:2012.08534  [astro-ph.CO]}.


\bibitem{TT3}
G.~E.~Addison, Y.~Huang, D.~J.~Watts, C.~L.~Bennett, M.~Halpern, G.~Hinshaw, 
J.~L.~Weiland,
``Quantifying discordance in the 2015 Planck CMB spectrum,''
{\hypersetup{urlcolor=vividviolet}\href{https://iopscience.iop.org/article/10.3847/0004-637X/818/2/132}{ApJ   \textbf{818} (2016) 132}}, \href{https://arxiv.org/abs/1511.00055}{arXiv:1511.00055  [astro-ph.CO]}.






\bibitem{rsd}
N.~Kaiser,
``Clustering in real space and in redshift space,''
{\hypersetup{urlcolor=vividviolet}\href{https://academic.oup.com/mnras/article/227/1/1/1065830?login=false}{Mon. Not. Roy. Astron. Soc. \textbf{227} (1987) 1}}.


\bibitem{tension1}
E.~Macaulay, I.~K.~Wehus, H.~K.~Eriksen,
``A Lower Growth Rate from Recent Redshift Space Distortion Measurements than Expected from Planck,''
{\hypersetup{urlcolor=vividviolet}\href{https://journals.aps.org/prl/abstract/10.1103/PhysRevLett.111.161301}{Phys. Rev. Lett. \textbf{111} (2013) 161301}}, \href{https://arxiv.org/abs/1303.6583}{arXiv:1303.6583 [astro-ph.CO]}.

\bibitem{tension2}
{\it Planck Collaboration},
``Planck 2015 results. XIV. Dark energy and modified gravity,''
{\hypersetup{urlcolor=vividviolet}\href{https://www.aanda.org/articles/aa/full_html/2016/10/aa25814-15/aa25814-15.html
}{A\&A \textbf{594} (2016) A14}}, \href{https://arxiv.org/abs/1502.01590}{arXiv:1502.01590[astro-ph.CO]}.

\bibitem{tension3}
S.~Dodelson, S.~Park,
``Nonlocal Gravity and Structure in the Universe,''
{\hypersetup{urlcolor=vividviolet}\href{https://journals.aps.org/prd/abstract/10.1103/PhysRevD.90.043535}{Phys. Rev. D \textbf{90} (2014) 043535}}, \href{https://arxiv.org/abs/1310.4329}{arXiv:1310.4329 [astro-ph.CO]}; {\it Erratum} {\hypersetup{urlcolor=vividviolet}\href{https://journals.aps.org/prd/abstract/10.1103/PhysRevD.98.029904}{Phys. Rev. D \textbf{98} (2018) 029904}}.

\bibitem{intertwined3}
E.~Di Valentino, L.~A.~Anchordoqui, O.~Akarsu, Y.~Ali-Haimoud, L.~Amendola, N.~Arendse, M.~Asgari, M.~Ballardini, S.~Basilakos and E.~Battistelli, \textit{et al.}
``Cosmology intertwined III: $f\sigma_8$ and $S_8$,''
{\hypersetup{urlcolor=vividviolet}\href{https://www.sciencedirect.com/science/article/pii/S0927650521000487}{Astropart. Phys. \textbf{131}, 102604 (2021)
}}, \href{https://arxiv.org/abs/2008.11285}{arXiv:2008.11285 [astro-ph.CO]}.



\bibitem{cosmot1}
S.~Weinberg,  {\it Cosmology},  (Oxford University Press, Oxford, 2008). 

\bibitem{cosmot2}
C.~Catto\"{e}n, M.~Visser,
``The Hubble series: Convergence properties and redshift variables,''
{\hypersetup{urlcolor=vividviolet}\href{https://iopscience.iop.org/article/10.1088/0264-9381/24/23/018}{Class. Quantum  Grav.  \textbf{24} (2007) 5985}}, \href{https://arxiv.org/abs/0710.1887}{arXiv:0710.1887 [gr-qc]}.

\bibitem{snap}
M.~Dunajski, G.~Gibbons,
``Cosmic Jerk, Snap and Beyond,''
{\hypersetup{urlcolor=vividviolet}\href{https://iopscience.iop.org/article/10.1088/0264-9381/25/23/235012}{Class. Quantum  Grav.  \textbf{25} (2008) 235012}}, \href{https://arxiv.org/abs/0807.0207}{arXiv:0807.0207 [gr-qc]}.


\bibitem{state}
V.~Sahni, T.~D.~Saini, A.~A.~Starobinsky, U.~Alam,
``Statefinder -- a new geometrical diagnostic of dark energy,''
{\hypersetup{urlcolor=vividviolet}\href{https://link.springer.com/article/10.1134/1.1574831
}{JETP Lett.  \textbf{77} (2003) 201}}, \href{https://arxiv.org/abs/astro-ph/0201498}{arXiv:0201498 [astro-ph]}.



\bibitem{reza1}
A.~Mehrabi, M.~Rezaei,
``Cosmographic Parameters in Model-independent Approaches,''
{\hypersetup{urlcolor=vividviolet}\href{https://iopscience.iop.org/article/10.3847/1538-4357/ac2fff}{ApJ \textbf{923} (2021) 274}}, \href{https://arxiv.org/abs/2110.14950}{arXiv:2110.14950 [astro-ph.CO]}.


\bibitem{reza2}
P.~Mukherjee,  N.~Banerjee,
``Non-parametric reconstruction of the cosmological \textit{jerk} parameter,''
{\hypersetup{urlcolor=vividviolet}\href{https://link.springer.com/article/10.1140/epjc/s10052-021-08830-5
}{Eur. Phys. Jour. C \textbf{81} (2021) 36}}, \href{https://arxiv.org/abs/2007.10124}{arXiv:2007.10124 [astro-ph.CO]}.


\bibitem{intj3}
A.~Mukherjee,  N.~Banerjee,
``Parametric reconstruction of the cosmological jerk from diverse observational data sets,''
{\hypersetup{urlcolor=vividviolet}\href{https://journals.aps.org/prd/abstract/10.1103/PhysRevD.93.043002}{Phys. Rev. D \textbf{93}  (2016)  043002}}, \href{https://arxiv.org/abs/1601.05172}{arXiv:1601.05172 [gr-qc]}.


\bibitem{intj1}
Z.~-X.~Zhai, M.~-J.~Zhang, Z.~-S.~Zhang, X.~-M.~Liu, T.~-J.~Zhang,
``Reconstruction and constraining of the jerk parameter from OHD and SNe Ia observations,''
{\hypersetup{urlcolor=vividviolet}\href{https://www.sciencedirect.com/science/article/pii/S0370269313008137?via%3Dihub
}{Phys. Lett. B \textbf{727} (2013) 8}}, \href{https://arxiv.org/abs/1303.1620}{arXiv:1303.1620 [astro-ph.CO]}.

\bibitem{intj2}
H.~Amirhashchi, S.~Amirhashchi,
``Recovering $\Lambda$CDM model from a cosmographic study,''
{\hypersetup{urlcolor=vividviolet}\href{https://link.springer.com/article/10.1007/s10714-020-2664-5
}{Gen. Rel. Grav.  \textbf{52} (2020) 13}}, \href{https://arxiv.org/abs/1811.05400}{arXiv:1811.05400 [astro-ph.CO]}.

\bibitem{cosmoII}
A.~Mukherjee,  N.~Banerjee,
``In search of the dark matter dark energy interaction: a kinematic approach,''
{\hypersetup{urlcolor=vividviolet}\href{https://iopscience.iop.org/article/10.1088/1361-6382/aa54c8}{Class. Quantum Grav. \textbf{34} (2017) 035016}}, \href{https://arxiv.org/abs/1610.04419}{arXiv:1610.04419 [astro-ph.CO]}.

\bibitem{scalefactor}
S.~G.~Choudhury, A.~Dasgupta, N.~Banerjee,
``Reconstruction of $f(R)$ gravity models for an accelerated universe using Raychaudhuri equation,''
{\hypersetup{urlcolor=vividviolet}\href{https://academic.oup.com/mnras/article/485/4/5693/5380788?login=false
}{Mon. Not. Roy. Astron. Soc. \textbf{485} (2019) 5693}}, \href{https://arxiv.org/abs/1903.04775}{arXiv:1903.04775 [gr-qc]}.	


\bibitem{recoQ1}
T.~Yang, Z.~-K.~Guo, R.~-G.~Cai,
``Reconstructing the interaction between dark energy and dark matter using Gaussian Processes,''
{\hypersetup{urlcolor=vividviolet}\href{https://journals.aps.org/prd/abstract/10.1103/PhysRevD.91.123533}{Phys. Rev. D \textbf{91} (2015)  123533}}, \href{https://arxiv.org/abs/1505.04443}{arXiv:1505.04443 [astro-ph.CO]}.

\bibitem{recoQ2}
P.~Mukherjee, N.~Banerjee,
``Nonparametric reconstruction of interaction in the cosmic dark sectors,''
{\hypersetup{urlcolor=vividviolet}\href{https://journals.aps.org/prd/abstract/10.1103/PhysRevD.103.123530}{Phys. Rev. D \textbf{103} (2021)  123530}}, \href{https://arxiv.org/abs/2105.09995}{arXiv:2105.09995 [astro-ph.CO]}.


\bibitem{braun}
D.~Pavon,  B.~Wang, ``Le Chatelier-Braun principle in cosmological physics,''
{\hypersetup{urlcolor=vividviolet}\href{https://www.doi.org/10.1007/s10714-008-0656-y}{Gen. Rel. Grav. {\bf 41} (2009) 1}},  \href{https://arxiv.org/abs/0712.0565}{arXiv:0712.0565 [gr-qc]}.

\bibitem{brauncc}
P.~Mukherjee,  N.~Banerjee, ``Non-parametric reconstruction of interaction in the cosmic dark sector,''
{\hypersetup{urlcolor=vividviolet}\href{https://journals.aps.org/prd/abstract/10.1103/PhysRevD.103.123530}{Phys. Rev. D {\bf 103} (2021)  123530}},  \href{https://arxiv.org/abs/2105.09995}{arXiv:2105.09995 [astro-ph.CO]}.

\bibitem{salvatelli}
V.~Salvatelli, N.~Said, M.~Bruni, A.~Melchiorri,  D.~Wands, ``Indications of a late-time interaction in the dark sector,"
{\hypersetup{urlcolor=vividviolet}\href{https://journals.aps.org/prl/abstract/10.1103/PhysRevLett.113.181301}{Phys. Rev. Lett.  {\bf 113} (2014) 181301}},  \href{https://arxiv.org/abs/1406.7297}{arXiv:1406.7297 [astro-ph.CO]}.	



\bibitem{coincidence}
H.~E.~S.~Velten, R.~vom Marttens,  W.~Zimdahl, ``Aspects of the cosmological ``coincidence problem"," {\hypersetup{urlcolor=vividviolet}\href{https://link.springer.com/article/10.1140%2Fepjc%2Fs10052-014-3160-4}{Eur. Phys. Jour. C \textbf{74} (2014)  3160}}, \href{https://arxiv.org/abs/arXiv:1410.2509}{arXiv:1410.2509 [astro-ph.CO]}.


\bibitem{fluid1}
V.~F.~Cardone,   C.~Tortora, A.~Troisi,  S.~Capozziello, ``Beyond the perfect fluid hypothesis for the dark energy equation of state,"
{\hypersetup{urlcolor=vividviolet}\href{https://journals.aps.org/prd/abstract/10.1103/PhysRevD.73.043508}{Phys. Rev. D \textbf{73} (2006)   043508}},
\href{https://arxiv.org/abs/astro-ph/0511528}{arXiv:0511528 [astro-ph]}.



\bibitem{fluid2}
P.~-H.~Chavanis, ``The Logotropic Dark Fluid as a unification of dark matter and dark energy,"  \href{https://www.sciencedirect.com/science/article/pii/S0370269316301150?via%3Dihub}{\color{vividviolet}{Phys. Lett. B \textbf{758} (2016) 59}}, \href{https://arxiv.org/abs/1505.00034}{arXiv:1505.00034 [astro-ph.CO]}.


\bibitem{fluid3}
S.~Capozziello, R.~D'Agostino, R.~Giamb\'o, O.~Luongo, ``Effective field description of the Anton-Schmidt cosmic fluid,"
{\hypersetup{urlcolor=vividviolet}\href{https://journals.aps.org/prd/abstract/10.1103/PhysRevD.99.023532}{Phys. Rev. D  \textbf{99} (2019)   023532}}, 
		\href{https://arxiv.org/abs/1810.05844}{arXiv:1810.05844 [gr-qc]}.


\bibitem{fluid4}
S.~D.~Odintsov,  V.~K.~Oikonomou,  A.~V.~Timoshkin, E.~N.~Saridakis, R.~Myrzakulov, ``Cosmological Fluids with Logarithmic Equation of State,"
{\hypersetup{urlcolor=vividviolet}\href{https://www.sciencedirect.com/science/article/abs/pii/S0003491618302604?via%3Dihub}{Ann. of Phys.  \textbf{398} (2018)  238}}, 
\href{https://arxiv.org/abs/1810.01276v1}{arXiv:1810.01276 [gr-qc]}.


\bibitem{fluid5}
D.~Bini, A.~Geralico, D.~Gregoris, S.~Succi, ``Dark energy from cosmological fluids obeying a Shan-Chen non-ideal equation of state,"
{\hypersetup{urlcolor=vividviolet}\href{https://journals.aps.org/prd/abstract/10.1103/PhysRevD.88.063007}{ Phys.  Rev. D  \textbf{88} (2013)  063007}}, 
\href{https://arxiv.org/abs/1408.5483}{arXiv:1408.5483 [gr-qc]}.

\bibitem{cg}
A.~Y.~Kamenshchik, U.~Moschella,  V.~Pasquier,
``An Alternative to quintessence,''
{\hypersetup{urlcolor=vividviolet}\href{https://www.sciencedirect.com/science/article/pii/S0370269301005718}{Phys. Lett. B \textbf{511} (2001) 265}}, 
\href{https://arxiv.org/abs/gr-qc/0103004}{arXiv:0103004 [gr-qc]}.

\bibitem{gcg}
M.~C.~Bento, O.~Bertolami, A.~A.~Sen,
``Generalized Chaplygin gas, accelerated expansion and dark energy matter unification,''
{\hypersetup{urlcolor=vividviolet}\href{https://journals.aps.org/prd/abstract/10.1103/PhysRevD.66.043507}{Phys. Rev. D \textbf{66} (2002) 043507}}, 
\href{https://arxiv.org/abs/gr-qc/0202064}{arXiv:0202064 [gr-qc]}.


\bibitem{tamanini}
S.~Bahamonde, C.~G.~Boehmer, S.~Carloni, E.~J.~Copeland, W.~Fang,  N.~Tamanini,
``Dynamical systems applied to cosmology: dark energy and modified gravity,'' \href{https://www.sciencedirect.com/science/article/abs/pii/S0370157318302242?via%3Dihub}{\color{vividviolet}{Phys. Rep. \textbf{775} (2018)  1}}, \href{https://arxiv.org/abs/1712.03107}{arXiv:1712.03107 [gr-qc]}.

\bibitem{copeland}
E.~J.~Copeland, A.~R.~Liddle, D.~Wands,
``Exponential potentials and cosmological scaling solutions,'' \href{https://journals.aps.org/prd/abstract/10.1103/PhysRevD.57.4686}{\color{vividviolet}{Phys. Rev. D \textbf{57} (1998)  4686}}, \href{https://arxiv.org/abs/gr-qc/9711068}{arXiv:9711068 [gr-qc]}.


\bibitem{potentials1}
J.~J.~Halliwell, ``Scalar fields in cosmology with an exponential potential," {\hypersetup{urlcolor=vividviolet}\href{https://www.sciencedirect.com/science/article/abs/pii/0370269387910112}{Phys.  Lett. B \textbf{185} (1987)  341}}.

\bibitem{potentials2}
I.~P.~C.~Heard, D.~Wands, ``Cosmology with positive and negative exponential potentials," {\hypersetup{urlcolor=vividviolet}\href{https://iopscience.iop.org/article/10.1088/0264-9381/19/21/309}{Class. Quantum Grav. \textbf{19} (2002)  5435}}, \href{https://arxiv.org/abs/gr-qc/0206085v1}{arXiv:0206085 [gr-qc]}.

\bibitem{potentials3}
C.~Wetterich, ``An asymptotically vanishing time-dependent cosmological \lq\lq constant"," {\hypersetup{urlcolor=vividviolet}\href{https://adsabs.harvard.edu/full/1995A%26A...301..321W}{Astron. Astrophys. \textbf{301} (1995)  321}}, \href{https://arxiv.org/abs/hep-th/9408025v1}{arXiv:9408025 [hep-th]}.



\bibitem{potentials5}
P.~G.~Ferreira, M.~Joyce, ``Cosmology with a Primordial Scaling Field," {\hypersetup{urlcolor=vividviolet}\href{https://journals.aps.org/prd/abstract/10.1103/PhysRevD.58.023503}{Phys. Rev. D \textbf{58} (1998)  023503}}, \href{https://arxiv.org/abs/astro-ph/9711102v3}{arXiv:9711102 [astro-ph]}.

\bibitem{potentials6}
E.~V.~Linder, ``Pole Dark Energy," {\hypersetup{urlcolor=vividviolet}\href{https://journals.aps.org/prd/abstract/10.1103/PhysRevD.101.023506}{Phys. Rev. D \textbf{101} (2020)  023506}}, \href{https://arxiv.org/abs/1911.01606v1}{arXiv:1911.01606 [astro-ph.CO]}.

\bibitem{potentials7}
P.~J.~E.~Peebles, B.~Ratra, ``Cosmology with a Time-Variable Cosmological ``Constant''," {\hypersetup{urlcolor=vividviolet}\href{https://articles.adsabs.harvard.edu/pdf/1988ApJ...325L..17P}{Astrophys. J. Lett. \textbf{325} (1988)  L17}}.


\bibitem{chimento}
L.~P.~Chimento, ``Linear and nonlinear interactions in the dark sector," {\hypersetup{urlcolor=vividviolet}\href{https://journals.aps.org/prd/abstract/10.1103/PhysRevD.81.043525}{Phys. Rev. D \textbf{81} (2010)  043525}}, \href{https://arxiv.org/abs/0911.5687v2}{arXiv:0911.5687 [astro-ph.CO]}.


\bibitem{coleybook}
A.~Coley, {\it Dynamical Systems and Cosmology},
(Springer, The Netherlands, 2003).






\bibitem{cmb1}
J.~R.~Bond,  G.~Efstathiou,  M.~Tegmark, ``Forecasting Cosmic Parameter Errors from Microwave Background Anisotropy Experiments," {\hypersetup{urlcolor=vividviolet}\href{https://adsabs.harvard.edu/full/1997MNRAS.291L..33B}{Mon. Not. Roy. Astron. Soc. \textbf{291} (1997)  L33}}, \href{https://arxiv.org/abs/astro-ph/9702100}{arXiv:9702100 [astro-ph]}.

\bibitem{cmb2}
M.~Zaldarriaga, D.~Spergel, U.~Seljak, ``Microwave Background Constraints on Cosmological Parameters," {\hypersetup{urlcolor=vividviolet}\href{https://iopscience.iop.org/article/10.1086/304692}{ApJ \textbf{488} (1997)  1}}, \href{https://arxiv.org/abs/astro-ph/9702157}{arXiv:9702157 [astro-ph]}.

\bibitem{cmb3}
G.~Efstathiou, J.~R.~Bond, ``Cosmic Confusion: Degeneracies among Cosmological Parameters Derived from Measurements of Microwave Background Anisotropies," {\hypersetup{urlcolor=vividviolet}\href{https://academic.oup.com/mnras/article/304/1/75/972361?login=false}{Mon. Not. Roy. Astron. Soc. \textbf{304} (1999)  75}}, \href{https://arxiv.org/abs/astro-ph/9807103}{arXiv:9807103 [astro-ph]}.


\bibitem{sunny1}
S.~Vagnozzi, A.~Loeb, M.~Moresco, ``Eppur \`{e} piatto? The cosmic chronometer take on spatial curvature and cosmic concordance," {\hypersetup{urlcolor=vividviolet}\href{https://iopscience.iop.org/article/10.3847/1538-4357/abd4df}{ApJ \textbf{908} (2021)  84}}, \href{https://arxiv.org/abs/2011.11645}{arXiv:2011.11645 [astro-ph.CO]}.


\bibitem{sunny2}
S.~Dhawan, J.~Alsing, S.~Vagnozzi, ``Non-parametric spatial curvature inference using late-Universe cosmological probes," {\hypersetup{urlcolor=vividviolet}\href{https://academic.oup.com/mnrasl/article-abstract/506/1/L1/6292268?redirectedFrom=fulltext&login=false}{Mon. Not. Roy. Astron. Soc. \textbf{506} (2021)  L1}}, \href{https://arxiv.org/abs/2104.02485}{arXiv:2104.02485 [astro-ph.CO]}.

\bibitem{prior}
G.~Efstathiou, ``To $H_0$ or not to $H_0$?," {\hypersetup{urlcolor=vividviolet}\href{https://academic.oup.com/mnras/article/505/3/3866/6293858?login=false}{Mon. Not. Roy. Astron. Soc. \textbf{505} (2021)  3866}}, \href{https://arxiv.org/abs/2103.08723}{arXiv:2103.08723 [astro-ph.CO]}.


\bibitem{nature1}
G.~-B.~Zhao, M.~Raveri, L.~Pogosian, Y.~Wang, R.~G.~Crittenden, W.~J.~Handley, W.~J.~Percival, F.~Beutler, J.~Brinkmann, C.~-H.~Chuang, A.~J.~Cuesta, D.~ J.~ Eisenstein, F.~-S.~Kitaura, K.~Koyama, B.~L'Huillier, R.~C.~Nichol, M.~M.~ Pieri, S.~Rodriguez-Torres, A.~J.~Ross, G.~Rossi, A.~G.~Sanchez, A.~ Shafieloo, J.~L.~Tinker, R.~Tojeiro, J.~A.~Vazquez, H.~Zhang, ``Dynamical dark energy in light of the latest observations," {\hypersetup{urlcolor=vividviolet}\href{https://www.nature.com/articles/s41550-017-0216-z}{Nature Astronomy \textbf{1} (2017)  627}}, \href{https://arxiv.org/abs/1701.08165}{arXiv:1701.08165 [astro-ph.CO]}.


\bibitem{diva}
E.~Di Valentino, A.~Melchiorri, O.~Mena, ``Can interacting dark energy solve the $H_0$ tension?," {\hypersetup{urlcolor=vividviolet}\href{https://journals.aps.org/prd/abstract/10.1103/PhysRevD.96.043503}{Phys. Rev. D \textbf{96} (2017)   043503}}, \href{https://arxiv.org/abs/1704.08342}{arXiv:1704.08342 [astro-ph.CO]}.


\bibitem{oin}
B.~-H.~Lee, W.~Lee, E.~\'O Colg\'ain, M.~M.~Sheikh-Jabbari, S.~Thakur, ``Is local $H_0$ at odds with dark energy EFT?," {\hypersetup{urlcolor=vividviolet}\href{https://iopscience.iop.org/article/10.1088/1475-7516/2022/04/004}{JCAP \textbf{04} (2022)   004}}, \href{https://arxiv.org/abs/2202.03906}{arXiv:2202.03906 [astro-ph.CO]}.

\bibitem{future1}
A.~G.~Riess, W.~Yuan, S.~Casertano, L.~M.~Macri,  D.~Scolnic, ``“The Accuracy of the Hubble Constant Measurement Verified through Cepheid Amplitude," {\hypersetup{urlcolor=vividviolet}\href{https://iopscience.iop.org/article/10.3847/2041-8213/ab9900}{ApJ Lett. \textbf{896} (2020)  L43}}, \href{https://arxiv.org/abs/2005.02445}{arXiv:2005.02445 [astro-ph.CO]}.

\bibitem{future2}
S.~Birrer,  T.~Treu, ``DCOSMO - V. Strategies for precise and accurate measurements of the Hubble constant with
strong lensing," {\hypersetup{urlcolor=vividviolet}\href{https://www.aanda.org/articles/aa/full_html/2021/05/aa39179-20/aa39179-20.html}{A\&A \textbf{649} (2021)  A61}}, \href{https://arxiv.org/abs/2008.06157}{arXiv:2008.06157 [astro-ph.CO]}.


\bibitem{future3}
{\it EUCLID Collaboration}, ``Euclid Definition Study Report," \href{https://arxiv.org/abs/1110.3193}{arXiv:1110.3193 [astro-ph.CO]}.


\bibitem{future4}
{\it DESI Collaboration}, ``The DESI Experiment Part I: Science,Targeting, and Survey Design,"  \href{https://arxiv.org/abs/1611.00036}{arXiv:1611.00036 [astro-ph.IM]}.


\bibitem{f(R)1}
P.~K.~S.~Dunsby, E.~Elizalde, R.~Goswami, S.~Odintsov, D.~Saez-Gomez,
``On the LCDM Universe in f(R) gravity,''
{\hypersetup{urlcolor=vividviolet}\href{https://journals.aps.org/prd/abstract/10.1103/PhysRevD.82.023519}{Phys. Rev. D \textbf{82} (2010)  023519}}, \href{https://arxiv.org/abs/1005.2205}{arXiv:1005.2205 [gr-qc]}.



\bibitem{f(R)2}
J.~-h.~He, B.~Wang,
``Revisiting $f(R)$ gravity models that reproduce $\Lambda$CDM expansion,''
{\hypersetup{urlcolor=vividviolet}\href{https://journals.aps.org/prd/abstract/10.1103/PhysRevD.87.023508}{Phys. Rev. D \textbf{87} (2013)  023508}}, \href{https://arxiv.org/abs/1208.1388}{arXiv:1208.1388 [astro-ph.CO]}.



\bibitem{f(R)3}
S.~Fay, S.~Nesseris, L.~Perivolaropoulos,
``Can f(R) Modified Gravity Theories Mimic a LCDM Cosmology?,''
{\hypersetup{urlcolor=vividviolet}\href{https://journals.aps.org/prd/abstract/10.1103/PhysRevD.76.063504}{Phys. Rev. D \textbf{76} (2007)  063504}}, \href{https://arxiv.org/abs/gr-qc/0703006}{arXiv:0703006 [gr-qc]}.








\bibitem{f(R)5}
M.~Ortiz-Ba\~nos, M.~Bouhmadi-L\'opez, R.~Lazkoz, V.~Salzano,
``${\Lambda}$CDM suitably embedded in f(R) with a non-minimal coupling to matter,''
{\hypersetup{urlcolor=vividviolet}\href{https://link.springer.com/article/10.1140/epjc/s10052-021-09004-z}{Eur. Phys. J. C \textbf{81} (2021)  237}}, \href{https://arxiv.org/abs/2103.01982}{arXiv:2103.01982 [gr-qc]}.



\bibitem{Chakraborty}
S.~Chakraborty, K.~MacDevette, P.~Dunsby,
``A model independent approach to the study of $f(R)$ cosmologies with expansion histories close to $\Lambda$CDM,''
{\hypersetup{urlcolor=vividviolet}\href{https://www.doi.org/10.1103/PhysRevD.103.124040}{Phys. Rev. D \textbf{103} (2021)  124040}}, \href{https://arxiv.org/abs/2103.02274}{arXiv:2103.02274 [gr-qc]}.	



\bibitem{ltb}
P.~Dunsby, N.~Goheer, B.~Osano, J.~-P.~Uzan,
``How close can an Inhomogeneous Universe mimic the Concordance Model?,''
{\hypersetup{urlcolor=vividviolet}\href{https://iopscience.iop.org/article/10.1088/1475-7516/2010/06/017}{JCAP \textbf{06} (2010)  017}}, \href{https://arxiv.org/abs/1002.2397}{arXiv:1002.2397 [astro-ph.CO]}.





\end{thebibliography}
\end{document}